\documentclass[fleqn, usenatbib]{mnras}

\usepackage{newtxtext,newtxmath}

\usepackage[T1]{fontenc}

\usepackage{graphicx}	
\usepackage{amsmath}	
\usepackage{mathrsfs}
\usepackage[flushleft]{threeparttable}
\usepackage{siunitx}
\usepackage{booktabs, caption}
\usepackage{float}
\usepackage{pdflscape}
\usepackage{ulem}
\usepackage{physics}

\newcommand{\GG}[1]{}

\title[Peeling Back Layers of Extinction]{Peeling Back the Layers of Extinction of Dusty Galaxies in the Era of JWST: Modelling Joint NIRSpec + MIRI Spectra at rest-frame 1.5 - 28 $\mu$m}

\author[F. R. Donnan et al.]{
F. R. Donnan, $^{1}$\thanks{E-mail: fergus.donnan@physics.ox.ac.uk} I. Garc{\'i}a-Bernete,$^{1}$ D. Rigopoulou,$^{1, 2}$ M. Pereira-Santaella,$^{3}$ P. F. Roche,$^{1}$ A. Alonso-Herrero$^{4}$ \\
\\
$^{1}$Department of Physics, University of Oxford, Keble Road, Oxford, OX1 3RH, UK\\
$^{2}$School of Sciences, European University Cyprus, Diogenes street, Engomi, 1516 Nicosia, Cyprus\\
$^{3}$ Instituto de F\'{\i}sica Fundamental, CSIC, Calle Serrano 123, E-28006, Madrid, Spain\\
$^{4}$ Centro de Astrobiolog\'{\i}a (CAB), CSIC-INTA, Camino Bajo del Castillo s/n, E-28692 Villanueva de la Ca\~nada, Madrid, Spain
}

\date{Accepted XXX. Received YYY; in original form ZZZ}

\pubyear{2015}

\begin{document}
\label{firstpage}
\pagerange{\pageref{firstpage}--\pageref{lastpage}}
\maketitle

\begin{abstract}
We present an analysis of the combined NIRSpec and MIRI spectra of dusty galaxies between 1.5 - 28 $\mu$m restframe by implementing a differential extinction model, where the strength of extinction varies across the spectrum as different layers of the obscuring dust are probed. Our model is able to recover a 2D distribution of dust temperature and extinction which allows inference of the physical nature of the dust in these environments. We show that differential extinction is necessary to reproduce the spectra of 4 highly obscured Luminous Infrared Galaxies observed with NIRSpec IFU and MIRI MRS, where simple screen or uniformly mixed dust distributions fail to fit the data. We additionally compare the extinction of HII regions in these galaxies via hydrogen recombination lines, the extinction of molecular gas via the H$_2$ lines, Polycyclic Aromatic Hydrocarbons via the 12.7/11.3 PAH ratio and the stellar continuum. We find that the molecular gas is deeply buried with the HII regions in star-forming regions, with a similar extinction to the hottest dust components. However we find the cooler dust to be less obscured, at a similar extinction to the stellar continuum and PAHs. The nuclei show a complex dust distribution with VV114 NE, NGC 3256 S, IIZw96 SW showing a deeply buried continuum source relative to the molecular gas/HII regions. Additionally, NGC 3256 S, NGC 7469 and VV114 SW show an isolated hot dust component, indicative of AGN heating, where NGC 3256 S and NGC 7469 are previously known AGN.

\end{abstract}




\section{Introduction}

Luminous Infrared Galaxies (LIRGs, L\textsubscript{IR} $>10^{11}  \rm{L}_{\odot}$) and Ultraluminous Infrared Galaxies (ULIRGs, L\textsubscript{IR} $>10^{12}  \rm{L}_{\odot}$) are dust and gas rich galaxies, with dust enshrouded star formation and AGN activity \citep[e.g.][]{Sanders1996, Rigopoulou1999, Lonsdale2006}. These galaxies play a key role in the evolution of galaxies, making up the bulk of the star formation rate density at cosmic noon \citep[e.g.][]{LeFloch2005, Magnelli2011}. In addition $\sim 40\%$ of local ULIRGs contain Compact Obscured Nuclei \citep[CONs;][]{Falstad2021, Garcia-Bernete2022, Donnan2023}, which hide rapid, active supermassive black hole growth and/or extremely compact starbursts \citep[e.g.][]{Veilleux2009, Aalto2015, Aalto2019, Pereira-Santaella2021}.

As a consequence of the vast quantities of dust, emission in the optical is absorbed and reprocessed in the infrared. Therefore observations in the infrared are required to accurately measure properties of these galaxies such as the star-formation rate, properties of the ISM etc. In particular, the mid-infrared is a feature rich region of the spectrum, with dust emission and absorption signatures, Polycyclic Aromatic Hydrocarbons (PAHs), and numerous emission lines from ionised and molecular gas.

With all these emission/absorption features, the mid-IR spectra of galaxies can be extremely complex, therefore extracting the emission features and determining accurately the level of the continuum emission is a challenge. As the PAH features are broad and blended they create a pseudo-continuum, under which the true continuum lies \citep[e.g.][]{Brandl2006, Smith2007, Gallimore2010, Li2020, Rigopoulou2021}. 
This means a simple, local continuum around emission features is often not sufficient to obtain accurate PAH fluxes, particularly around the 7.7 $\mu$m band. This problem is compounded in more obscured sources, where strong absorption features from water ice ($\sim 6 \mu$m), CH ($\sim 7 \mu$m) and silicates ($\sim 9.8 \mu$m) create a complex continuum \citep[e.g.][]{Spoon2007, IDEOS, Veilleux2009, Garcia-Bernete2023}.

In highly obscured environments it becomes difficult to reproduce the continuum, particularly over a large wavelength range. This may be a consequence of differential extinction, where multiple regions of different dust densities all contribute within a single aperture. The resulting spectrum is therefore a combination of different intrinsic spectra with varying levels of obscuration. Such environments have been mapped in the Milky Way \citep[e.g.][]{Lallement2022}. Differential extinction has an impact on the shape of the observed continuum, such as changing the ratio of the 18 $\mu$m to 9.8$\mu$m silicate features. This has been observed in AGN tori with clumpy torus models showing a higher 18 $\mu$m to 9.8 $\mu$m ratio than smooth torus models \citep[e.g.][]{Hatziminaoglou2015, Martinez-Peredes2020, Garcia-Bernete2023}.

Different approaches have been used to model the mid-IR spectra of galaxies, however the majority of these assume a single level of extinction i.e an extinction law is scaled by a single extinction value over the whole spectrum. Popular choices include \textsc{PAHFIT} \citep{Smith2007}, which models the full spectrum with a series of blackbodies subject to extinction and a series of PAH emission features, with flexible profiles. This works well for relatively unobscured star-forming galaxies which are dominated by PAH emission but can struggle to fit AGN and/or obscured galaxies \citep[e.g.][]{Gallimore2010}. Another code, \textsc{CAFE} \citep{Marshall2007}, similarly uses modified blackbodies but is restricted to only three where each has a different extinction.

If an AGN is present, one approach is to model the dust from AGN using a template, typically produced from radiative transfer simulations of the torus \citep[e.g.][]{RamosAlmeida2009, Alonso-Herrero2011, Alonso-Herrero2012, Martinez-Paredes2015, Martinez-Peredes2020, Martiniz-Paredes2021, Gonzalez-Martin2019a,Gonzalez-Martin2019, Garcia-Bernete2019, Garcia-Bernete2022d}. This approach is valuable in providing constrains on physical parameters of the torus such as opening angle, inclination etc. However as this is often highly model dependent, one often needs to fit multiple libraries to obtain a complete picture of the torus for a given object. A wide variety of libraries exist in the literature, with different geometries, dust distributions etc. These include smooth torus models \citep[e.g.][]{Fritz2006, Efstathiou2013}, clumpy models \citep[e.g.][]{Nenkova2008a, Nenkova2008b, Honig2010}, two-phase models \citep[e.g.][]{Stalevski2016} and models incorporating dust outflows \citep[e.g.][]{Honig2017}. These templates can also struggle to fit the data in certain cases such as for the most obscured nuclei, where the deep silicate feature is difficult to reproduce \citep{Garcia-Bernete2022}, with only the most obscured smooth tori able to generate the deep silicates. As a result, semi-empirical templates of highly obscured galaxies have often been used to reproduce their spectra such as that employed in QUESTFIT \citep[][]{Veilleux2009}, however these often fail to fit the data due to a lack of model flexibility.

While the techniques outlined above are extremely valuable, there is a need for a more general, data driven approach that can fit a wide range of objects such as star-forming regions, type 1/2 AGN, extremely obscured nuclei, etc, but still provide inference of interesting/useful physical parameters of the dusty structure in these galaxies such as the degree of obscuration and dust temperature. This is especially true in the era of JWST where the much higher data quality of NIRSpec/MIRI \citep[e.g.][]{Wells2015, Wright2015, Rieke2015, Jakobsen2022, Argyriou2023, Boker2023} over previous instruments such as Spitzer or Akari, means the current models can struggle to fit the data due to an increase in spectral resolution and sensitivity. Additionally, the significant improvement in spatial resolution enables better isolation of the nuclei of galaxies, and therefore the spectra are less diluted by star-formation. As a consequence, the spectra are more dominated by a complex dust continuum, requiring new models to reproduce.

In this work we present a new model, where we describe the continuum with a generalised differential extinction model to fit combined NIRSpec and MIRI spectra from 1.5 $\mu$m to 28 $\mu$m, where dust of different temperatures are subject to different levels of extinction. The structure of this paper is as follows. We first describe the model in detail in Section \ref{sec:Model}. We then test with simulated data in Section \ref{sec:SimData} to verify the model before applying it to JWST data in Section \ref{sec:JWSTData}. Subsequently we compare the dust extinction model with other tracers of extinction, namely the stellar continuum, Hydrogen recombination lines and H$_2$ lines for the 4 LIRGs observed with NIRSpec and MIRI and discuss the inferred dusty structure in Section \ref{sec:Discuss}.

\section{Model}
\label{sec:Model}
The simplest way to predict the observed spectrum of a galaxy, $f_{\nu}$, is to model the intrinsic spectrum, $f_{\nu}^{\textrm{Int}}$, with a sum of modified blackbodies in the infrared, which is then subject to a single screen of extinction of the form 
\begin{equation}
\label{eqn:Screen}
    \frac{f_{\nu}}{f_{\nu}^{\textrm{Int}}} = e^{ -\tau_{9.8} \tau(\lambda)}
\end{equation}
where $\tau(\lambda)$ is an assumed extinction law that is normalised to 1 at 9.8 $\mu$m. The level of extinction is then controlled by the optical depth at 9.8 $\mu$m, $\tau_{9.8}$, which scales the extinction law to increase the extinction. This is equivalent to $A_v$, which is often used for rest-frame UV/optical data. For reference $\tau_{9.8} \sim 0.1 A_v $ \citep[e.g.][]{Draine1989}. While mathematically simple, the screen geometry is not well motivated, as the obscuring dust and source dust are assumed to be separate, where the obscuring dust does not to emit (see Fig. \ref{fig:Cartoons}).

Alternatively, one can assume a mixed geometry where the emitting and obscuring dust is uniformly mixed \citep[e.g.][]{Smith2007}, which has the form 
\begin{equation}
\label{eqn:Mixed}
    \frac{f_{\nu}}{f_{\nu}^{\textrm{Int}}} = \frac{1-e^{ -\tau_{9.8} \tau(\lambda)  }}{\tau_{9.8} \tau(\lambda)}. 
\end{equation}
This is more physically motivated for star-forming regions compared to a simple screen, as demonstrated in Fig. \ref{fig:Cartoons}. 
These simple approaches tend to fail however when modelling highly obscured galaxies, particularly over a long wavelength range, where the emergent spectrum at $<5$ $\mu$m may be from a much less obscured environment (with a different intrinsic spectrum) compared to longer wavelength emission. A simple solution is to model two components, a relatively unobscured star-forming component and an obscured nucleus, such as that implemented in PAHDecomp \citep[][]{Donnan2023}. This is also the approach of \citet{Hernan-Caballero2016} and references therein. This however does not work for sources where there may be a gradient of extinction as we probe different layers as a function of wavelength. An example where such a geometry may be present is shown in \ref{fig:Cartoons}. Therefore in order to be able to fit the spectrum of any (U)LIRG, we require a more general model, to capture this ``differential extinction'' behaviour. This is described in detail in Section \ref{sec:DiffExt}. 

In addition to the differential extinction model, we model the PAH emission features following a similar procedure to that of \textsc{PAHFIT} \citep{Smith2007} and include additional molecular absorption features \citep[e.g.][]{Imanishi2010, Lai2020, Lai2023}. Our full model, $f_{\nu}$, is given by 
\begin{equation}
\label{eqn:Model}
    f_{\nu} = \sum_{i=1}^{N_{\rm PAH}} I_{\nu, \rm PAH}^{(i)} (\lambda) + C_{\nu} e^{ -\tau_{\rm ices (Dust)} (\lambda)} + S_{\nu} e^{ -\tau_{\rm ices (Stellar)} (\lambda)   } 
\end{equation}
where the first term is a sum over $N_{\rm PAH}$ PAH profiles $I_{\nu, \rm PAH} (\lambda)$ and the second is the total continuum comprising of the dust model incorporating differential extinction, $C_{\nu}$, and a stellar continuum, $S_{\nu}$, both of which are subject to a screen of the following absorption features. For the dust continuum we include absorption from water ice at $\sim$ 3 $\mu$m and 6$\mu$m, CO$_2$ at 4.0 $\mu$m and CH at $\sim$ 7$\mu$m while for the stellar continuum we only apply the near-IR absorption features of water ice at $\sim$ 3 $\mu$m and CO$_2$ at 4.0 $\mu$m. These are described in detail in the following sections. 

\begin{figure*}
	\includegraphics[width=\textwidth]{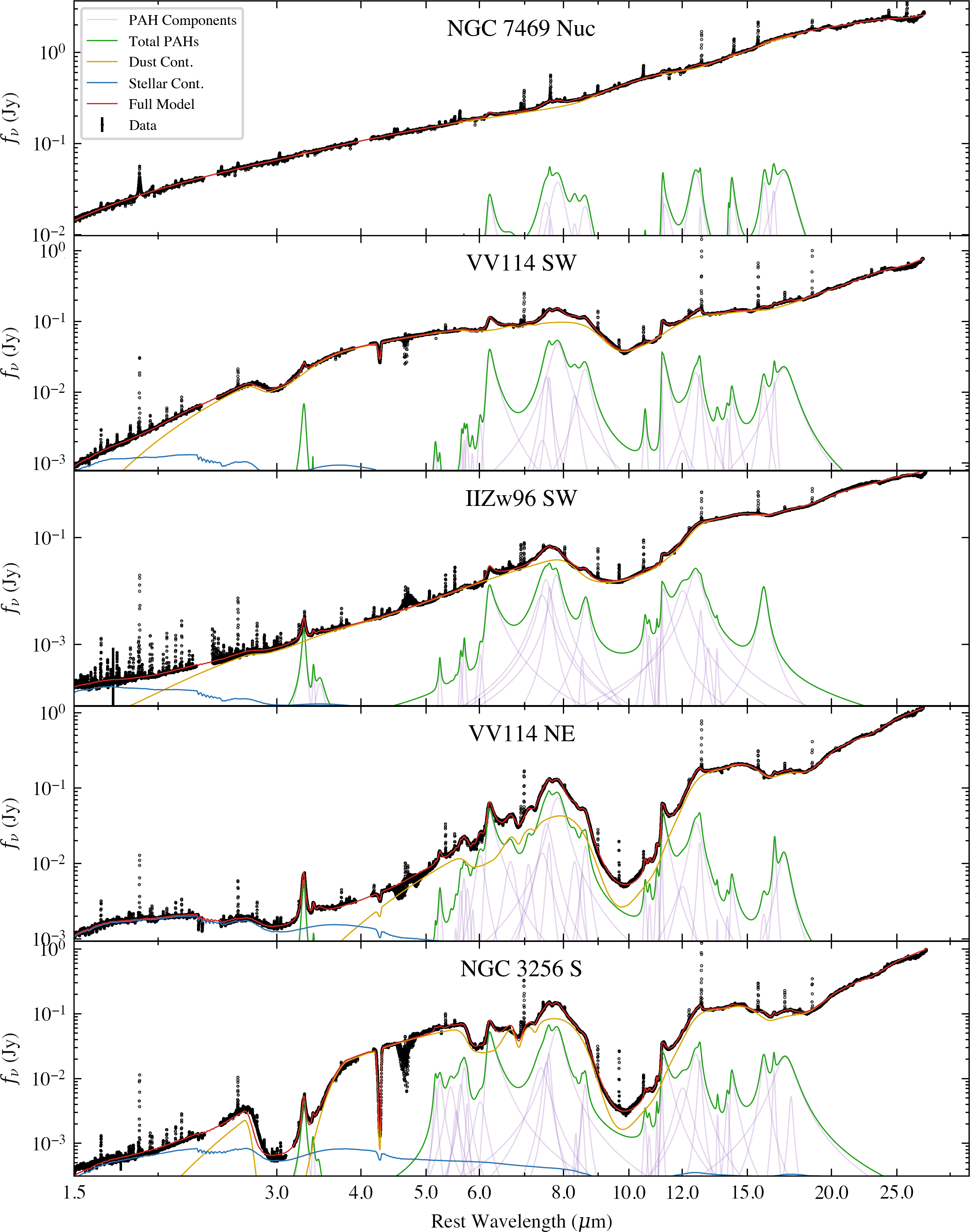}
    \caption{Five examples of fits to combined NIRSpec and MIRI spectra for different types of nuclei. From top to bottom is NGC 7469 (type 1 AGN - see Section \ref{sec:NGC7469Dis}), VV114 SW (Potential Type 2 AGN \citep{Rich2023} - see Section \ref{sec:VV114Dis}), IIZw96 SW (Obscured AGN \citep{Garcia-Bernete2023b} - see Section \ref{sec:IIZw96Dis}), VV114 NE (Obscured AGN \citep{Donnan2023b} - see Section \ref{sec:VV114Dis}) and NGC 3256 S  (Type 2 (Compton Thick) AGN \citep{Ohyama2015} - see Section \ref{sec:NGC3256Dis}). Note the CO band at $\sim$ 4.6 $\mu$m is masked in the fitting.}
    \label{fig:PAHComps} 
\end{figure*}

We show five examples of fits with this model to a variety of spectra in Fig. \ref{fig:PAHComps} to demonstrate the ability to fit a range of different objects with varying obscuration. The components comprising the model in equation \ref{eqn:Model} are shown.

To fit this model to data we use a Bayesian approach, where we maximise the following log probability 
\begin{equation}
    \label{eqn:Prob}
    \ln \textrm{Prob} = -\chi^2 - \sum \ln\sigma^2 - \Gamma P + \textrm{const.},
\end{equation}
where $\chi^2$ is simply the sum of the squared normalised residuals between the model and the data, $\sigma$ is the data error bars, and $P$ is a penalty for complex models of the dust distribution, scaled by a regularisation scale, $\Gamma$. This factor controls the level of regularisation, where a larger value increases the penalty for non-smooth dust distributions. This term is described in detail in Section \ref{sec:Regularisation}.

\subsection{Continuum}
\subsubsection{Differential Extinction}
\label{sec:DiffExt}
A general model of dust emission and extinction can be written as a 2D weighted average of modified blackbodies, $\epsilon_{\nu} B_{\nu}$, and screens of extinction, $e^{ -\tau_{9.8} \tau(\lambda)  }$, where the weights are parameterised by a 2D distribution, $\Psi(T, \tau_{9.8})$, of dust temperature, $T$, and extinction, $\tau_{9.8}$. The modified blackbodies are given by the product of the Planck function, $B(T)$, and a wavelength dependent emissivity term, $\epsilon_{\nu}$ (see Section \ref{sec:Emissivity}). The dust continuum is therefore given by

\begin{equation}
\label{eqn:Continuum}
    C_{\nu} (\lambda) = A \int_{0}^{\infty}\int_{35 \textrm{K}}^{1500 \textrm{K}} \Psi \left ( T, \tau_{9.8} \right ) e^{ -\tau_{9.8} \tau(\lambda)  } \frac{\epsilon_{\nu} B_{\nu}(T, \lambda)}{B_0(T)} dT d\tau_{9.8}, 
\end{equation}
where the sum of the weights, $\Psi(T, \tau_{9.8})$, is unity and $B_0(T)$ is a normalisation factor for each modified blackbody. We choose $B_0(T)$ to be the integrated flux between 1.5$\mu$m and 28$\mu$m with some emissivity, $\epsilon_{\nu}$. We do not consider temperatures below 35K as dust this cold will not contribute significantly to the flux below 30 $\mu$m. We use a maximum temperature of 1500K as this is approximately the average dust sublimation temperature for a mix of silicate and graphite grains \citep[e.g.][]{Kishimoto2007}.

In this work we assume a fixed extinction law, $\tau(\lambda)$, using that presented in \citet{Donnan2023b}, which consists of a power law and absorption from silicates at $\sim9.8$ and $\sim18$ microns. We choose this law as the silicate features include absorption from crystalline features, necessary for highly obscured spectra \citep[e.g.][]{IDEOS, Donnan2023b}. As the weighted average describes only the shape of the model, the entire continuum is scaled by a factor $A$ to match the data.

In the simplest case where there is a single screen of extinction, the dust distribution has a form 
\begin{equation}
    \Psi(T, \tau^{'}_{9.8}) = \rho(T) \delta\left( \tau^{'}_{9.8} - \tau_{9.8}\right), 
\end{equation}
where $\rho(T)$ describes the distribution of blackbodies of different temperatures (see Fig. \ref{fig:Cartoons}). The single screen is represented by a Dirac delta function which simplifies the integral into 
\begin{multline}
    C_{\nu} (\lambda) = e^{ -\tau_{9.8} \tau(\lambda)  } \int_{35 \textrm{K}}^{1500 \textrm{K}} \rho(T) \frac{\epsilon_{\nu} B_{\nu}(T, \lambda)}{B_0(T)} dT
\end{multline}
where the observed continuum, $C_{\nu} (\lambda)$, is the intrinsic multiplied by a factor of $e^{ -\tau_{9.8} \tau(\lambda)}$, recovering equation (\ref{eqn:Screen}). 

For the mixed case there is a uniform dust distribution between 0 and some $\tau_{9.8}$ (see Fig. \ref{fig:Cartoons}):
\begin{equation}
    \Psi(T, \tau^{'}_{9.8}) = \rho(T) \begin{cases} 
      \frac{1}{\tau_{9.8}} & 0\leq \tau^{'}_{9.8} \leq \tau_{9.8} \\
      0 & \textrm{otherwise}
   \end{cases}
\end{equation}
which simplifies the integral to 
\begin{equation}
    C_{\nu} (\lambda) = \frac{1-e^{ -\tau_{9.8} \tau(\lambda)  }}{\tau_{9.8} \tau(\lambda)} \int_{35 \textrm{K}}^{1500 \textrm{K}} \rho(T) \frac{\epsilon_{\nu} B_{\nu}(T, \lambda)}{B_0(T)} dT, 
\end{equation}
where the observed continuum is the intrinsic continuum multiplied by a factor $ \frac{1-e^{ -\tau_{9.8} \tau(\lambda)  }}{\tau_{9.8} \tau(\lambda)}$, recovering equation (\ref{eqn:Mixed}). This is the original \textsc{PAHFIT} model \citep{Smith2007}, where the amplitude of different blackbodies is allowed to vary, comprising $\rho(T)$, and is then subject to the extinction factor.

To implement our differential extinction model, instead of fixing the dust distribution, $\Psi \left ( T, \tau_{9.8} \right )$, we constrain this from the data by fitting the continuum with a flexible model for $\Psi \left ( T, \tau_{9.8} \right )$. We do this under some prior restrictions, to prevent nonphysical solutions for $\Psi \left ( T, \tau_{9.8} \right )$, by ensuring the distribution is smooth and only applies differential extinction where necessary to fit the data.

To implement this model one could choose a parametric form of $\Psi \left ( T, \tau_{9.8} \right )$ such as a 2D Gaussian or a 2D Lorentzian, however there is no obvious/physically motivated choice for the form of this distribution. Assuming a given parametric form may therefore bias the results. We therefore opt for a non-parametric approach where we bin $\Psi \left ( T, \tau_{9.8} \right )$ into a 20x20 grid and allow each grid element to vary individually. This approach is similar to the stellar population modelling of \textsc{pPXF} \citep{Cappellari2022}, where a non-parametric 2D distribution of stellar metalicity and age is used to generate the model continuum. 

We choose the grid carefully to ensure that each grid element gives a similar weight to the continuum. We bin $\tau_{9.8}$ into 20 evenly spaced values in natural log space between, $\tau_{9.8} = 0.05$ and $\tau_{9.8} = 15$. By binning in $\ln$ space, we are evenly sampling different screens of extinction of the form $e^{ -\tau_{9.8} \tau(\lambda)  } $. For the temperature axis we bin in $\log_{10}$ space between 35K and 1500K. To ensure each grid element has equal weighting when fitting, we normalise the spectrum of each grid element by the integrated flux over the wavelength range of the data. After fitting, we re-scale each grid element such that the normalisation matches that in equation \ref{eqn:Continuum} to obtain the $\Psi \left ( T, \tau_{9.8} \right )$ in that equation.

The wavelength dependent extinction factor is given by the ratio of the observed to the intrinsic continuum such as those shown in equation (\ref{eqn:Screen}) and (\ref{eqn:Mixed}) for the screen and mixed cases respectively. From our model, the intrinsic continuum is defined as
\begin{equation}
    f_{\nu}^{\textrm{Int}} = S_{\nu}^{\textrm{Int}} + A \int_{0}^{\infty}\int_{35 \textrm{K}}^{1500 \textrm{K}} \Psi \left ( T, \tau_{9.8} \right ) \frac{\epsilon_{\nu} B_{\nu}(T, \lambda)}{B_0(T)} dT d\tau_{9.8}, 
\end{equation}
where $S_{\nu}^{\textrm{Int}}$ is the stellar population model defined in equation (\ref{eqn:Stellar}). From this we calculate the wavelength dependent extinction factor
\begin{equation}
\label{eqn:ExtCor}
    \frac{f_{\nu}}{f_{\nu}^{\textrm{Int}}} = \frac{S_{\nu} + C_{\nu}}{f_{\nu}^{\textrm{Int}}}
\end{equation}
where $S_{\nu}$ and $C_{\nu}$ are defined in equations (\ref{eqn:Continuum}) and (\ref{eqn:Stellar}) respectively while $f_{\nu}^{\textrm{Int}}$ is given above. In general $\frac{f_{\nu}}{f_{\nu}^{\textrm{Int}}} \neq e^{ -\tau_{9.8} \tau(\lambda) }$, as this only holds for a single screen geometry. Rather, the extinction factor depends on the inferred $\Psi \left ( T, \tau_{9.8} \right )$ and therefore can vary between objects. For example, if there is a buried hot dust component compared to the cool dust, the extinction factor will drop at $\sim 5$ $\mu$m compared to long wavelengths. This is demonstrated with simulated data in Section \ref{sec:SimData}.

\subsubsection{Regularisation}
\label{sec:Regularisation}
Allowing each grid element on a 20x20 grid to vary, effectively adds 400 parameters to the model (typical NIRSpec+MIRI spectra contain $\sim$ 20000 data points). This provides the flexibility to fit complex, obscured galaxy spectra, however the best fit solution for $\Psi \left ( T, \tau_{9.8} \right )$ may not be physically meaningful with so many parameters. While we find MCMC sampling returns reasonably smooth solutions compared to a simple maximisation of the $\ln \textrm{Prob}$ (equation (\ref{eqn:Prob})), we include some  regularisation to favour solutions where the distribution is simple/smooth.

We follow the approach of \citet{Cappellari2017, Cappellari2022} where we add a penalty factor to the log-likelihood to disfavour complex models of $\Psi \left ( T, \tau_{9.8} \right )$. In particular we use the sum of the Laplacian for each row and column of $\Psi$:
\begin{multline}
\label{eqn:Laplace}
    P =  \int_{0}^{\infty}\int_{35 \textrm{K}}^{1500 \textrm{K}} \Psi ''\left ( T, \tau_{9.8} \right ) dT d\tau_{9.8} =  \\
    \sum_{i} \left(\Psi_{i-1, j}  - 2 \Psi_{i, j} + \Psi_{i+1, j}\right)^2 + \sum_{j} \left(\Psi_{i, j-1}  - 2 \Psi_{i, j} + \Psi_{i, j+1}\right)^2
\end{multline}
where the integral of the 2\textsuperscript{nd} derivative of $\Psi$ is calculated numerically on the 20x20 grid in each axis as shown by the second line of the equation. This factor must scale with the number of data points to be comparable to the other terms in equation (\ref{eqn:Prob}). We therefore choose $\Gamma = 1000 N$, where there are $N$ data points in a given spectrum.

In addition to the preference for smooth solutions, we include additional regularisation on the ``effective'' $\tau_{9.8}$ as function of wavelength. During the fit, the effective $\tau_{9.8}$ is calculated for 20 wavelength bins across the spectrum, evenly spaced in $\log_{10}\lambda$. This is done by computing the extinction factor ($f_{\nu}^{\textrm{Obs}}/f_{\nu}^{\textrm{Intr}}$) at each wavelength bin and converting to an effective $\tau_{9.8}$ where $\tau_{9.8}^{\textrm{Eff}} = -\ln \left( f_{\nu}^{\textrm{Obs}}/f_{\nu}^{\textrm{Intr}} \right)$ at 9.8 $\mu$m. $\tau_{9.8}^{\textrm{Eff}}$ is then converted into units to match the grid elements of $\Psi \left ( T, \tau_{9.8} \right )$, such that we are smoothing in steps of $e^{-\tau_{9.8}}$. We then place a Gaussian prior on the differences between successive $\tau_{9.8}^{\textrm{Eff}}$ with a mean of zero and a standard deviation of 6 grid elements.  We chose this prior to be wide such as to not limit the inference on $\Psi \left ( T, \tau_{9.8} \right )$, but restrictive enough to prevent the model fitting for differential extinction where it is not needed. Therefore this prior ensures that the fit defaults to no differential extinction and only deviates from this when necessary to fit the data. 

\subsubsection{Emissivity}
\label{sec:Emissivity}
The shape of the blackbody profile for a given temperature, $T$, is modified by an emissivity factor, $\epsilon_{\nu}$. Typically this factor takes the form of a power law
\begin{equation}
    \epsilon_{\nu} \propto \lambda^{-\beta},
\end{equation}
with values of $\beta = 1.5 - 2$. Codes such as \textsc{PAHFIT} \citep{Smith2007}, adopt $\beta=2$, however this approach does not account for silicates, which contribute significantly in the mid-infrared \citep[e.g.][]{Li2001}. Adopting a more physically motivated emissivity that includes silicates (as well as graphite grains), results in a more realistic spectrum including intrinsic silicate emission at 9.8 $\mu$m and 18$\mu$m  \citep[e.g.][]{Li2001, Draine2007, Marshall2007}. However, silicate emission is mainly observed in type 1 AGN \citep[e.g.][]{Gallimore2010, Garcia-Bernete2017, Martinez-Peredes2020, Garcia-Bernete2022b}, where lines of sight of hot silicates, heated by the black hole accretion disk, are visible without a significant hot background source, requiring clumpy or low-inclination tori. In star-forming regions, the continuum is typically flat or shows moderate silicate absorption. To achieve such a continuum with intrinsic silicate emission requires either a cool dust temperature, where the silicates contribute less and/or extra extinction to compensate. The problem is therefore degenerate from an observational perspective, as the observed continua for star-forming regions can be reproduced by either the power law emissivity or a physically motivated emissivity.

\begin{figure}
\includegraphics[width=\columnwidth]{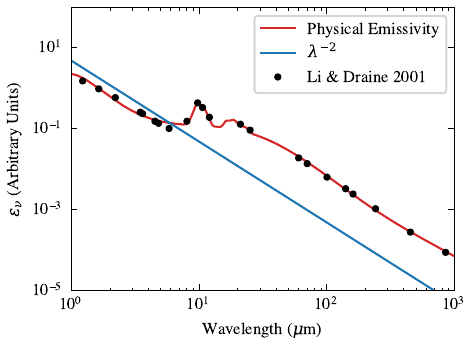}
\caption{Emissivity used to modify the intrinsic blackbody dust emission. The blue line shows a simple power law with index $\beta=2$. The black points are taken from Table 6 of \citet{Li2001} and include silicate $\&$ carbon grains. The red line shows our physical emissivity profile based on these points by fitting a polynomial and a silicate template.}
\label{fig:Emissivity} 
\end{figure}

In this work we explore both possibilities, starting with the simple power law with $\beta=2$ and then with a emissivity following the curve presented in Table 6 of \citet{Li2001}. To adjust to the spectral resolution of JWST, we fit these points with a polynomial and the empirical silicate templates in \citet{Donnan2023b}. This is shown in Fig. \ref{fig:Emissivity} in comparison with the simple power law. 

\subsection{Stellar Continuum}
In addition to the dust continuum, stellar emission will dominate below 3 $\mu$m. Fitting the stellar component is critical to constraining any possible contribution from hot dust, $T\sim 1000K$, at around $\sim$ 5 $\mu$m where the inclusion of NIRSpec to provide data at $<$ 5 $\mu$m provides strong constraints on the stellar contribution. Particular features of stellar continua can be seen such as CO absorption from stellar atmospheres at $\sim 2.2$ $\mu$m.

To model the stellar continuum we use two templates with ages of 10 Gyr and 100 Myr, at solar metallicity, provided by \textsc{FSPS} \citep{Conroy2009, Conroy2010}, with a \citet{Salpeter1995} Initial Mass Function (IMF). At wavelengths > 1 $\mu$m, the stellar continuum is dominated by more evolved stellar populations in LIRGs \citep[e.g.][]{Pereira-Santaella2015}, despite these galaxies being highly star-forming. We chose these two templates based on \citet{Pereira-Santaella2015}, as combination of these templates provides a reasonable stellar continuum without doing a full stellar population model, which would be beyond the scope of this work. We allow each template to be scaled individually to allow different combinations of young/older populations which relates to the star-formation history. The scaled sum of the templates is subject to its own screen of extinction, constrained by the shape of the spectrum at $\sim$ 1.5 - 2 $\mu$m. Therefore the stellar component is simply 
\begin{equation}
\label{eqn:Stellar}
    S_{\nu} = e^{ -\tau_{9.8} \tau(\lambda)  }S_{\nu}^{\textrm{Int}}, 
\end{equation}
where $S_{\nu}^{\textrm{Int}}$ is the sum of each template scaled by its corresponding scale factor.

\subsection{Ices}
In addition to the extinction present in the dust model, there are absorption features from H$_2$O ice, CO, CO$_2$ and aliphatic CH present in the spectrum as seen in equation (\ref{eqn:Model}). To account for this, we generate templates of the optical depth from highly obscured sources and apply them as a screen to the continuum. The templates for the mid-IR water ice and CH features at $\sim$ 6 $\mu$m are taken from \citet{Donnan2023}, which were generated from Spitzer IRS spectra of the deeply embedded source, NGC 4418 \citep[see also][]{Garcia-Bernete2023}. We find the low resolution Spitzer data is sufficient to generate a template for fitting the MIRI data as these features are broad. However, the near-IR water ice at $\sim$ 3 $\mu$m and CO$_2$ at $\sim$ 4.2 $\mu$m were generated using the NIRSpec spectrum of the southern nucleus of NGC 3256. This was necessary in order to create a reliable template as the depth of this feature, with the high signal to noise of NIRSpec, much better captures the shape compared to Akari data and the higher spatial resolution better isolates the nucleus.

\begin{figure}
\includegraphics[width=\columnwidth]{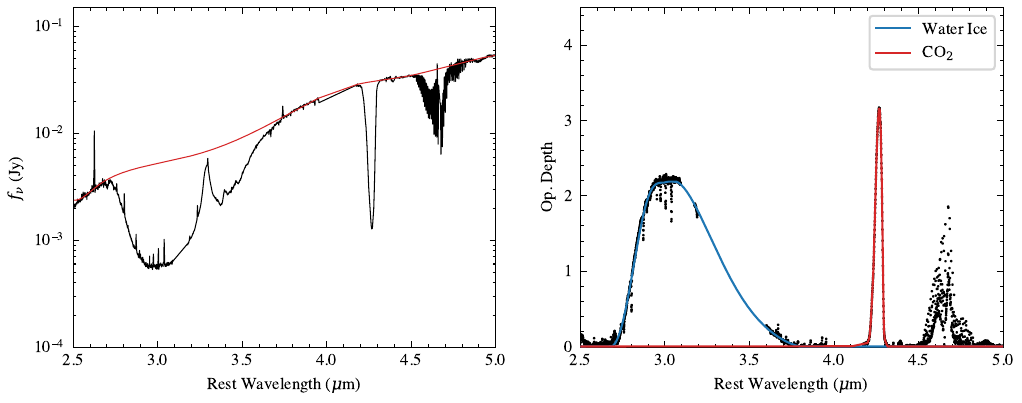}
\caption{Generation of water ice and CO$_2$ templates from the southern nucleus of NGC 3256. \textbf{Left:} NIRSpec spectrum between 2.5 and 5 $\mu$m shown in black. The red line shows a local spline continuum, masking the main absorption features of water ice, CO$_2$ and CO. \textbf{Right:} Optical depth curve of the absorption features using the local continuum in the left panel and masking the main emission features in the data. The blue line shows the generated template for water ice, created using two 5\textsuperscript{th} order polynomials for each side of the feature. The CO$_2$ template in red was generated by smoothing the data. We avoid generating a template for the CO feature and simply mask this region of the spectrum in the fitting.}
\label{fig:NIRIces} 
\end{figure}

We do this by first creating a local continuum using a cubic spline with anchor points at [2.5, 2.6, 2.65, 3.8, 4.18, 4.48, 4.82, 4.95] $\mu$m. This is shown in red in the left panel of Fig. \ref{fig:NIRIces}. We calculate an optical depth curve, $\tau_{\textrm{ices}} (\lambda)$, using this local continuum, $C_{\textrm{local}}$ where 
\begin{equation}
\label{eqn:Ice}
    \tau_{\textrm{ices}} (\lambda) = \ln \left( f_{\textrm{data}}/C_{\textrm{local}} \right)
\end{equation}
with $f_{\textrm{data}}$ as the spectra. We mask the main emission features before computing equation (\ref{eqn:Ice}), most notably the 3.3 $\mu$m and 3.4 $\mu$m PAH features. This results in the black points in the right panel of Fig. \ref{fig:NIRIces}. To convert this into a usable template we simply smooth the data for the CO$_2$ feature. For the broad water ice feature we fit two 5\textsuperscript{th} order polynomials to each side of the feature. 

We do not create a template for the CO features at $\sim$ 4.6 $\mu$m as this band consists of a variety of absorption/emission features which will change shape from object to object and thus requires more complex modeling \citep[e.g.][]{Pereira-Santaella2023, Gonzalez-Alfonso2023, Garcia-Bernete2023b, Buiten2023}. We therefore mask this region during the fitting process. However we find the extrapolated best fit continuum over this region provides a good fit to an underlying continuum (e.g. see Fig. \ref{fig:PAHComps}). 

We apply the near-infrared ices, namely the $\sim$ 3 $\mu$m water ice and CO$_2$ at $\sim$ 4.2 $\mu$m, to both the dust continuum and the stellar continuum. For both continua, the depth of the $\sim$ 3 $\mu$m water ice and the CO$_2$ feature are allowed to vary independently. Moreover, the depths of each feature are not necessarily the same for each continuum as a highly obscured AGN, such as NGC 3256 S, shows much deeper ices than typical stellar continua from star-forming regions. Therefore to obtain an adequate fit for NGC 3256 S, we indeed require different optical depths for the ices applied to the stellar vs dust continuum. This can be seen in the lower panel of Fig. \ref{fig:PAHComps}. We therefore allow the optical depth of the near-IR ices to vary separately as seen in equation (\ref{eqn:Model}). 

For most cases, there is not enough information in the data to constrain each component separately (unlike NGC 3256 S), and therefore we place a Gaussian prior that the depth of the ices is the same for the dust and stellar continuum, such that these are only separate components when necessary to fit the data. The prior is on the ratio of the two components with a mean of 1 and a standard deviation of 0.2, such that $\pm$5$\sigma$ varies between zero and two.

\subsection{PAH Profiles}
We model the PAH features with a series of Drude profiles, following the original approach of \citet{Smith2007}. As discussed in \citet{Donnan2023b}, we add additional profiles to be able to fit the higher spectral resolution data from JWST. To fit the NIRSpec portion of the spectra, we follow \citet{Lai2020}, where the original \textsc{PAHFIT} was extended to fit AKARI data.

To model asymmetric features, we introduce an asymmetry parameter to a single Drude profile to better capture the shape of the feature with fewer parameters than summing multiple components. This is applied to the 3.3, 3.4, 5.2, 6.2, 11.0, 11.3, 13.55 and 14.2 $\mu$m PAH profiles. We follow the approach of \citet{Gordon2021} and \citet{Stanick2008} where we introduce parameter, $a$, into the following equation for the  
\begin{equation}
\label{eqn:Drude}
    I_{\nu, \rm PAH} (\lambda) = A \frac{\left(\gamma/\lambda_0 \right)^2}{\left( \lambda/\lambda_0 - \lambda_0/\lambda\right)^2 + \left(\gamma/\lambda_0 \right)^2}
\end{equation}
where the original FWHM $\gamma_0$ is modified 
\begin{equation}
    \label{eqn:Asymm}
    \gamma = \frac{2\gamma_0}{1+e^{a(\lambda-\lambda_0)}}
\end{equation}
where $A$ is the amplitude and $\lambda_0$ is the central wavelength.

Similar to previous iterations of \textsc{PAHFIT}, we restrict the width, centre and now the asymmetry parameters to a small range of values to prevent additional degeneracies.

For the majority of the features the asymmetry is simply turned off. A full list of these parameters can be found in Table. \ref{tab:PAHs}. The individual Drude profiles can be seen in Fig. \ref{fig:PAHComps} for some example fits with the model.

We have additionally added some restrictions to the relative strengths of certain PAH features. In particular, the less prominent PAH features, such as those between the 6.2 $\mu$m and 7.7 $\mu$m cannot be too large relative to the 6.2 $\mu$m and 7.7 $\mu$m PAHs. This is to prevent these PAHs fitting the continuum when the PAH emission is weak. We therefore place the following restrictions on the integrated PAH flux between certain limits:
PAH[6.6, 7.3]/PAH[6.0, 6.5] < 1.0, PAH[6.6, 7.3]/PAH[7.4, 8.2] < 0.25, PAH[12.3, 13.0]/PAH[11.7, 12.2] < 2.8, PAH[11.7, 12.2]/PAH[11.2, 11.5] < 1.0, PAH[10.0, 11.1]/PAH[11.2, 11.5] < 0.4, where the values in the square brackets denote the integration limits. These were chosen based on the star-forming spectra where the PAH features are bright and well resolved.

\subsection{Emission Lines}
In the Spitzer era it was possible to model emission lines with simple Gaussian profiles \citep[e.g.][]{Smith2007}. This was necessary, not only to obtain fluxes for each line but to properly decompose lines that may be blended with PAH profiles. This was particularly important for the 12.7 $\mu$m PAH feature which was heavily blended with the strong [Ne II] line at 12.81 $\mu$m.

An early modified version of \textsc{PAHFIT} for MIRI/MRS data was presented in \citet{Donnan2023b} \citep[also used in][]{Garcia-Bernete2022c} which modelled each line with a 2 component profile to capture the more complex line profile observed with the significantly higher spectral resolution of MIRI/MRS (R $\sim$ 1500 - 3500 )  over Spitzer/IRS data (R $\sim$ 500) \citep[e.g.][]{Labiano2021}. Considering the abundance of lines now detected in typical MIRI spectra \citep[e.g.][]{Pereira-Santaella2022, Garcia-Bernete2022b, Vivian2022, Lai2022, Armus2023, Donnan2023b, Rich2023, Young2023}, and the number of parameters required per line to capture the profile, this approach is extremely inefficient. Fortunately, the high spectral resolution means that blending of emission lines and PAH features is no longer an issue as the narrow widths of the emission lines makes them easily distinguishable from the much broader PAH features. We therefore simply mask the emission lines where they are detected in the spectra, and integrate using the model subtracted data to obtain the corresponding line fluxes. 

\begin{figure}
	\includegraphics[width=\columnwidth]{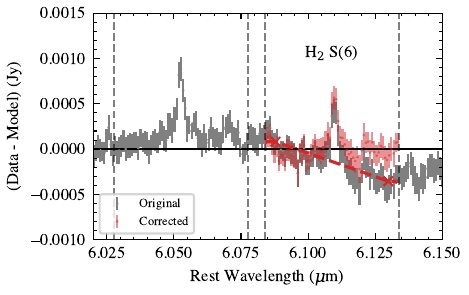}
    \caption{Model subtracted residuals for the fit to NGC 3256 SF1 (grey; see Section \ref{sec:DataReduction}), to be used to integrate to obtain emission line fluxes. The red dashed line marks a local continuum correction to the H$_2$ S(6) line, which when subtracted from the residuals, results in the red data points. This process is applied to every emission line before calculating the integrated flux. The vertical dashed lines show the range over which the line is integrated.}
    \label{fig:LineCorr} 
\end{figure}

For some emission lines, particular those with low signal to noise, the model subtracted data may contain some residuals which when integrated, will cause inaccurate flux measurements. This is particularly relevant if the line is faint. An example is shown in Fig. \ref{fig:LineCorr}, where we show the model subtracted residuals for two lines from a fit to a star-forming region in NGC 3256 (see Section \ref{sec:DataReduction}). The continuum around the  H$_2$ S(6) line (6.109 $\mu$m) appears overestimated resulting in negative residuals redwards of the line, whereas the continuum around first line \citep[H$_2$O $\nu$=1-0 212-101, 6.049 $\mu$m,][]{Gonzalez-Alfonso2023, Garcia-Bernete2023b} is predicted accurately. 

Before integrating to obtain a line flux, we correct the residuals by measuring a local continuum around the line and subtracting it from the residuals. We do this by calculating the median flux at either side of the line and connecting the points linearly. This creates the red dashed line in Fig. \ref{fig:LineCorr} which is subsequently subtracted to obtained the corrected residuals. This is done for every emission line before integrating to obtain a flux. We found that this correction led to underestimated flux errors as it effectively corrects for scatter between each MCMC sample. To overcome this, we re-sample the model subtracted residuals, for each MCMC sample, using the flux error bars before integrating to obtain a flux.

\section{Simulated Data}
\label{sec:SimData}
To assess the feasibility of this method in recovering the dust distribution from the data, we test the model with simulated data. In particular, we want to test that the model recovers accurately the input dust distribution. We also want to test that the model recovers the simple screen and mixed cases and thus does not fit for differential extinction where it does not exist.

\subsection{Generating Mock Data}
We generate three sets of mock data, a screen, mixed and a differential distribution where there is a gradient of obscuration, with the hot dust appearing more obscured than the cool dust. Specifically, we generated mock spectra by inputting a dust distribution $\Psi \left ( T, \tau_{9.8} \right )$ for each of the three cases. For the differential case we use three 2D Gaussian's with parameters $\left\{ x_0, y_0, \sigma_x, \sigma_y\right\} = \left\{ 4, 4, 2, 4 \right\} $, $\left\{ 8, 8, 2, 4\right\}$ and $\left\{ 12, 12, 2, 4\right\}$ in units of the grid elements with a position angle, of -40 $^{\circ}$. For these tests we adopt the simple power law emissivity profile (see Section \ref{sec:Emissivity}). This is to allow a more direct comparison to the traditional \textsc{PAHFIT} methods.

The dust distribution for the three sets of simulated data are shown in Fig. \ref{fig:SimData} with the corresponding continuum and extinction factor in the right panels. We add a stellar continuum to simulate the near-IR portion of the spectra. We use the 100 Myr template (we chose the 100 Myr arbitrarily here and test whether the model recovers the correct stellar continuum in Section \ref{sec:Case1}), and apply an extinction equal to that of the dust continuum for the screen and mixed cases. For the differential case, we apply a screen of extinction to the stellar continuum, equivalent to the screen case.

We additionally add a PAH and emission line template to the continuum before generating the mock data. This template was generated as the average of all the continuum subtracted spectra of the star-forming regions in NGC 3256 and NGC 7469 (see Section \ref{sec:DataReduction}) after fitting with the original \textsc{PAHFIT}-like continuum \citep[][]{Donnan2023b}. We then generate the mock data with wavelengths equal to that of the real data and noise of 1\% of the flux values. This gives a signal-to-noise comparable to the JWST data.

Each of the three simulated spectra are fitted with a fixed screen, fixed mixed and the new differential extinction model to investigate how well the true dust distribution and extinction factor can be recovered. The best fit extinction factor is shown in the right panels of Fig. \ref{fig:SimData}, in comparison with the true, input extinction factor. The contours in the left most panel show the best fit dust distribution from the differential model.
\subsection{Case 1: Screen Dust Distribution}
\label{sec:Case1}
For an input screen dust distribution, the screen model (see the top panel of Fig. \ref{fig:SimData}) recovers the true extinction factor within 1 $\%$ over the entire wavelength range as expected, while the mixed model performs the worst, underestimating the extinction across the wavelength range by $\sim 5 \%$ up to $\sim 15 \%$ at $\sim 4$ $\mu$m. The differential model recovers the true extinction well within 1 $\%$ across the majority of the wavelength range, only deviating at $\sim 4$ $\mu$m by a maximum of 4 $\%$. For all the fits we find the stellar continuum is recovered accurately, with the input stellar population of 100 Myr being recovered, as the near-IR is dominated by this component and so there is no difficulty in reproducing the input stellar continuum. 

The recovered dust distribution in the left panel of Fig. \ref{fig:SimData} peaks in the correct region but shows a larger spread. This larger spread is not seen for the mixed input data where the best fit dust distribution matches the spread accurately. Physically this is a bias towards more mixed dust distributions and so while this does not change the resultant extinction factor/continuum in any significant way, one must take care when drawing any inference about the physical dusty structure from the shape of $\Psi \left ( T, \tau_{9.8} \right )$. This bias towards more smooth dust distributions is a consequence of the regularisation with the non-parametric method. This is a known bias of non-parametric approaches \citep[e.g.][]{Leja2019, Iyer2019}, for example with non-parametric SFH-histories, where sudden bursts in star-formation can be physically expected but can difficult to reproduce with this approach.
\subsection{Case 2: Mixed Dust Distribution}

For the mixed input case we find the mixed model (see the middle panel of Fig. \ref{fig:SimData}) performs best as expected, recovering the extinction factor within $1\%$. The screen and differential models also perform well in this case, both finding the true extinction factor within 1$\%$ with both only deviating by a maximum of 4$\%$ at $\sim 1.6 \mu$m. These tests highlight that the differential extinction model is able to recover the extinction factor in the simple cases where there is no differential extinction. 

\subsection{Case 3: Differential Dust Distribution}

\begin{figure*}
	\includegraphics[width=\textwidth]{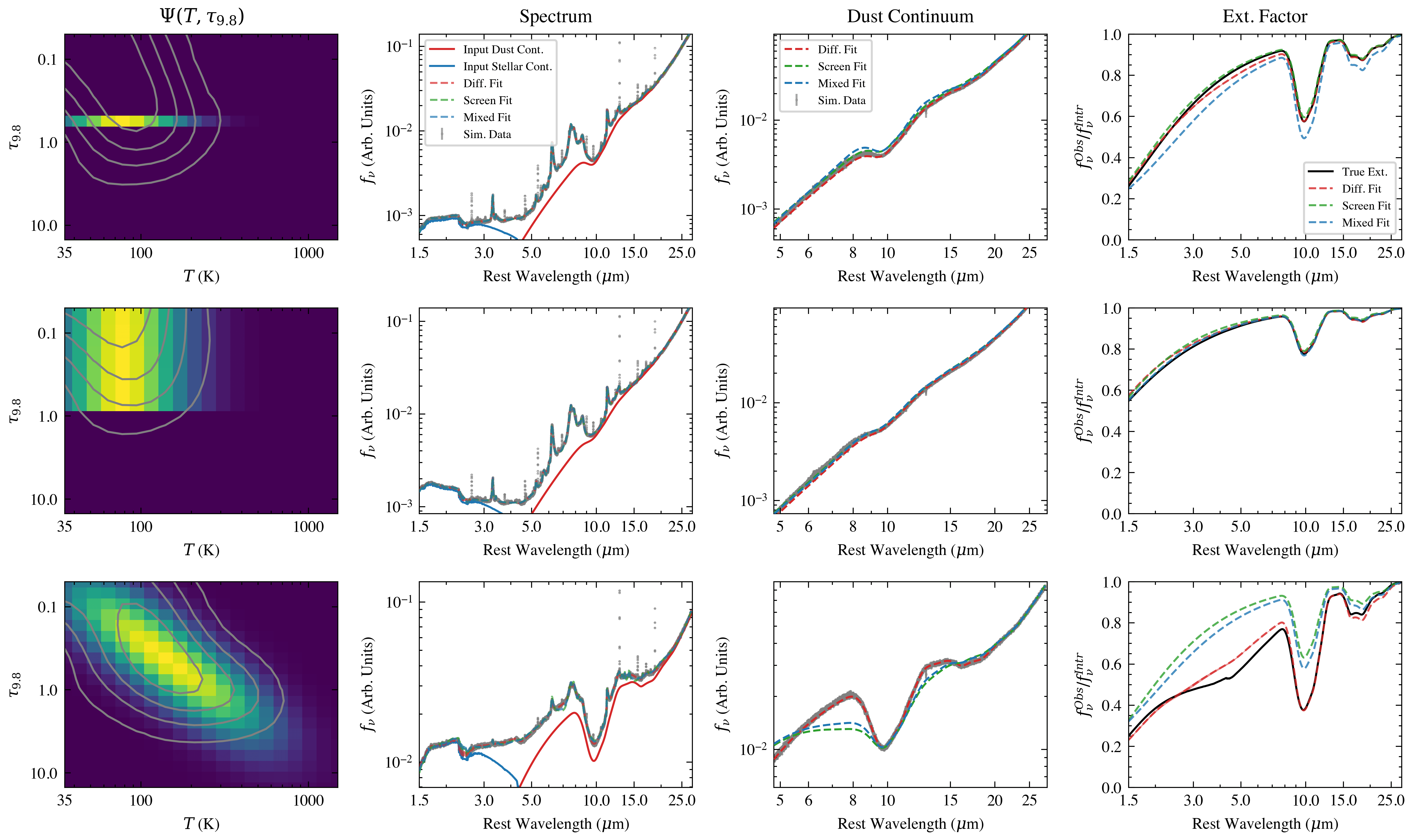}
    \caption{Tests with simulated data for three input dust distributions: screen (upper), mixed (middle) and differential (lower). The left panels show the dust distribution, $\Psi(T, \tau_{9.8})$, used to generate the simulated data overlayed with the best fit model in the grey contours. The colour scale is linear, with the grid elements summing to one. The 2\textsuperscript{nd} column shows the resulting continuum from the input dust distribution (solid red line) and the input stellar continuum (solid blue line). The simulated data is shown with the grey points.  The dashed lines show the different best fit models from the screen fit (green), mixed (blue) and the differential model (red). The 3\textsuperscript{rd} column shows just the dust continuum from the simulated data in comparison with the best fit models. The right panel shows the wavelength dependent extinction factor for the input (black) and the best fit results when fitting with a single screen (green dashed), pure mixed (blue dashed) and the new differential extinction model (red dashed). }
    \label{fig:SimData} 
\end{figure*}

The third simulated spectrum was produced using a dust distribution with a clear gradient where the hotter dust is more obscured than the colder dust (see the bottom panel of Fig. \ref{fig:SimData}). We find that the fixed screen and mixed models fail to recover the extinction factor, and indeed fail to produce a satisfactory fit to the continuum. The differential model however reproduces the input dust continuum accurately and recovers the extinction factor well, with a deviation at $\sim 5$ $\mu$m by up to 15$\%$. This deviation implies the model has difficulty recovering the most obscured hot dust, which would lower the extinction factor at $\sim 5$ $\mu$m. This is a limitation of the data as these obscured blackbodies will have a minimal contribution to the flux/shape of the model and thus makes little difference to the fit. Therefore due to the regularisation of the model, the fit favours more conservative solutions by not switching on components unnecessary to fit the data, pushing the recovered extinction factor closer to a case with less extreme differential extinction. This is therefore a hard limit of the data, where the information is simply not present to infer the most obscured hot dust that was included when generating the mock data.

For the third case we see that the traditional fixed or screen models do not properly fit the data and predict the extinction factor incorrectly, which would lead to incorrect flux ratios from emission features. This highlights a case where a more complex dust distribution is required and where our new model is able to reproduce the input conditions. As discussed in Section \ref{sec:Results}, dust distributions similar to this are present in LIRGs (and therefore also ULIRGs), and so accounting for the effect of differential extinction is necessary to model these spectra.

\section{JWST Data}
\label{sec:JWSTData}
\subsection{Targets}
\label{sec:Targets}
We use NIRSpec IFU and MIRI MRS data from Director's Discretionary Early Release Science Program 1328 (PI: Lee Armus \& Aaron Evans) of four local LIRGs, namely NGC 7469, NGC 3256, IIZw96 and VV 114. We show RGB NIRCAM images in Fig. \ref{fig:Apertures} for the 4 galaxies, with the F120W, F200W filters tracing stellar continuum and the F335M/F356W filters tracing the 3.3 $\mu$m PAH emission. The fully reduced files for the NIRCAM imaging were downloaded from the MAST archive.

NGC 7469 is a barred spiral galaxy hosting a type 1 AGN \citep[e.g.][]{Osterbrock1993} with a compact nuclear starburst and starburst ring \citep[e.g.][]{Diaz-Santos2007, Garcia-Bernete2022b, Lai2022, Lai2023}. The AGN drives a wide, highly ionised outflow towards the SE \citep[e.g.][]{Garcia-Bernete2022b, Vivian2022, 
Bianchin2023}, coinciding with a gap in the star-forming ring. This object has been subject to much analysis of the PAH population with JWST \citep[][]{Garcia-Bernete2022b, Zhang2023, Lai2023}.

NGC 3256 is a late-stage merger consisting of a face-on spiral galaxy \citep[NGC 3256 N][]{Sakamoto2014} with a bright nuclear starburst \citep{Lira2008}. The southern component of the merger (NGC 3256 S) is an edge on galaxy hosting a highly obscured, Compton Thick AGN \citep[e.g.][]{Ohyama2015} which drives a highly collimated outflow \citep[e.g.][]{Sakamoto2014, Pereira-Santaella2023}.

VV114 is a mid-stage major merger \citep[e.g.][]{Stierwalt2013} where the eastern component appears extremely dusty outputting a large infrared luminosity \citep{Armus2009}. The region is complex, with multiple nuclei and a large scale shock resulting from the interaction \citep[][]{Saito2017, Donnan2023b, Rich2011, Rich2023, Gonzalez-Alfonso2023, Buiten2023}. 

Finally, IIZw96 is also a major merger, with a dusty eastern region consisting of numerous compact, dusty star-forming clumps \citep{Inami2010, Inami2022}, separate from the two main nuclei of the interacting galaxies. The SW nucleus is thought to host a buried AGN from modelling IR rovibrational lines of the molecular bands \citep{Garcia-Bernete2023b}.

\subsection{Data Reduction}
\label{sec:DataReduction}

We reduced the data using modifications to the default pipeline.
The NIRSpec data were reduced using pipeline version 1.9.4 with the 1063 calibration context files. The details of the reduction can be found in \citet{Pereira-Santaella2023}.

The MIRI data were reduced using the default pipeline (v 1.9.4) up to the stage 2 data, using the 1041 calibration context. Before step 3 we run the residual fringe correction step offline. We run stage 3 without the background subtraction, instead performing this afterwards. We do this by calculating the median in each background frame before subtracting it. More details of the reduction are found in Appendix A of \citet{Garcia-Bernete2023}.

We extract spectra through a 0.3" radius aperture. To ensure the region is identical between the sub-channels we correct for small spatial misalignment's due to errors in the astrometry. We do this by measuring the position of the aperture we wish to extract the spectrum from relative to the brightest point source in the data. By measuring relative positions to a point source we avoid re-sampling the entire cube which may introduce further errors into the data. After extraction of the spectra, we run an additional residual fringe correction on the 1D spectra to remove the remaining fringing. Additionally, we use a PSF wavelength dependent correction factor for a point-like extraction for the nuclear apertures \citep[see also][]{Pereira-Santaella2022, Garcia-Bernete2022b, Donnan2023b}. This was obtained using a standard star for both NIRSpec and MIRI where the PSF correction factor is calculated as the ratio of the spectrum within 0.3" to the total spectrum as obtained from \textsc{CALSPEC} \citep{Bohlin2014, Bohlin2022}. For NIRSpec we use TYC 4433-1800-1 while for MIRI we use HD 163466.

To estimate errors for the fluxes at each wavelength we avoid those generated by the pipeline for the MIRI data as we found them to be underestimated. We instead use the dedicated background exposures and calculate the rms in each frame. For the NIRSpec data, we found the pipeline produced reasonable errors.

\begin{figure*}
	\includegraphics[width=\textwidth]{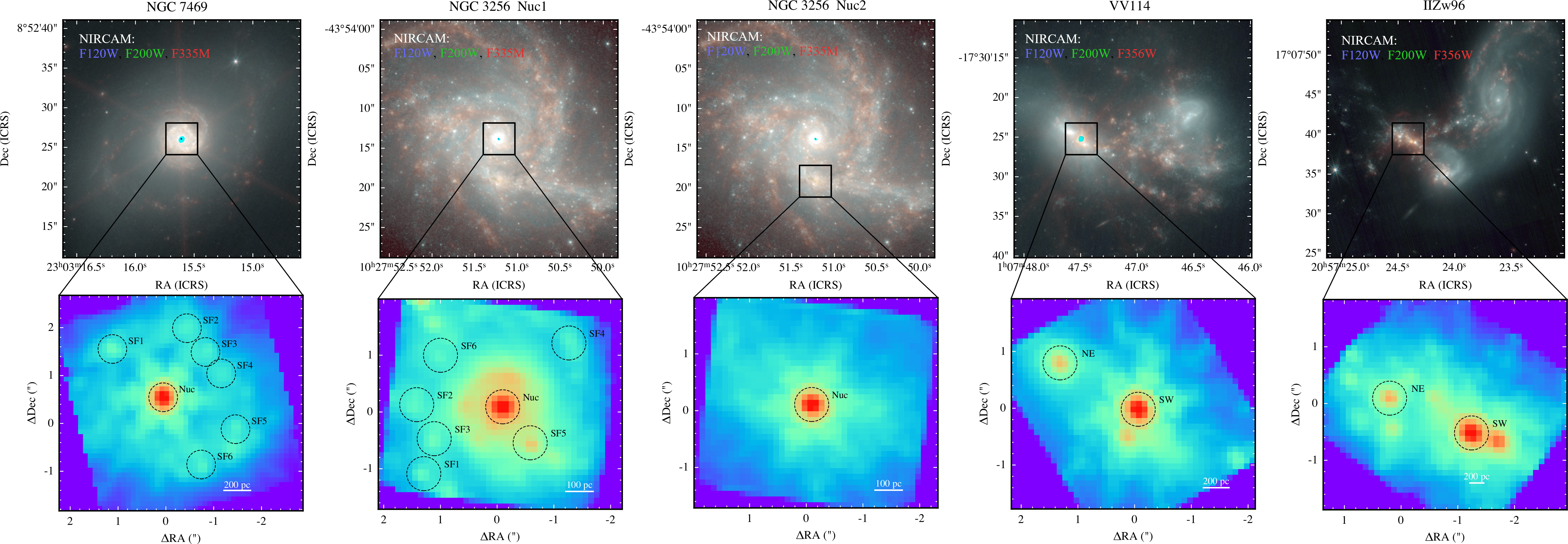}
    \caption{Images of the 4 LIRGs used in this work and the apertures used to extract the spectra. The upper panels show RGB images with NIRCAM. The F335M/F356W filters (red) show the 3.3 $\mu$m PAH emission while the F120W and F200W filters mainly trace stellar continuum. Most of the nuclei show some saturation at the brightest points. The field of view of the NIRSpec data is shown by the black box. The lower panels show the integrated NIRSpec detector 2 G395H/F290LP channel, integrated between 4.08 - 5.27 $\mu$m, tracing continuum (stellar and/or dust). The apertures used in this work to extract the spectra are shown and labelled. Each aperture has a radius of 0.3" and those labelled SF do not have any PSF correction applied as their emission appears extended. }
    \label{fig:Apertures} 
\end{figure*}

Fig. \ref{fig:Apertures} shows the positions of the apertures used in this work for each object. We extract six star-forming regions from each of NGC 7469 and NGC 3256, chosen based on near-IR bright regions. For NGC 7469, SF2, SF3 and SF4 lie on the redshifted region of the nuclear outflow \citep[e.g.][]{Garcia-Bernete2022b, Vivian2022, Bianchin2023} while a nuclear bar extends \citep[see Fig. 10 of ][]{Diaz-Santos2007} from SF1 to the southwest.

For IIZw96 SW, there is a secondary nucleus to the west, with a very small separation. At long wavelengths, the PSFs of each source overlap, leading to contamination. To mitigate this effect we extract the spectra of each nuclei, applying the aperture correction as discussed above. We then extract a total spectrum using a 0.6" aperture centred on the midpoint between the two nuclei. For this aperture we also apply a PSF correction factor by calculating the flux loss within a 0.6" aperture centred between two standard stars. We find the sum of the individual nuclei spectra to be larger than the 0.6" total spectrum, particularly at long wavelengths, due to contamination by each of the nuclei. To correct the SW nuclear spectrum, we subtract a fraction of the excess flux (where excess = sum of the nuclei - total), where the fraction is the contribution of the second nuclei to the sum of the nuclei. We find the correction to be small but significant at long wavelengths. For reference this is shown in Appendix \ref{app:IIZw96}.

\subsection{Other Estimates of Extinction}
\label{sec:EstofExt}
With the rich MIRI MRS and NIRSpec spectra, comes additional diagnostics of extinction. These can then be compared to the differential extinction model of the dust continuum to provide insight into the physical conditions of the regions. In the following sections we discuss three different measures of extinction using the emission features in the spectra.

\subsubsection{Hydrogen Recombination Lines (HI)}
\label{sec:HIExt}
Hydrogen recombination lines have been used in the optical to obtain extinction measurements for decades \citep[e.g.][]{Salim2020}. The so-called Balmer decrement measures the extinction of HII regions around young stars. This works by assuming an intrinsic H$\alpha$/H$\beta$ ratio under case B conditions \citep{Hummer1987} and inferring the optical depth required to reproduce the observed ratio. In the near and mid-infrared the same technique can be applied using the Brackett (Br, 4.05, 2.62, 2.17, ... $\mu$m), Pfund (Pf, 7.46, 4.65, 3.74, ... $\mu$m) and Humphreys (Hu, 12.4, 7.50, 5.91 ... $\mu$m) series rather than the Balmer series. 

While a large number of lines can be used, we focus on the ratio of Br$\beta$ (2.62 $\mu$m) to Br$\gamma$ (2.17 $\mu$m) as these lines have a high signal-to-noise and are unblended with any other lines.  A more detailed analysis combining all the HI lines may provide further insight but is beyond the scope of this work.

We predict the extinction for each spectrum by applying a screen with a given $\tau_{9.8}$ that reproduces the observed flux ratio. We do this for each MCMC sample to obtain uncertainties for the extinction.

We do not apply this to the nucleus of NGC 7469 as there is significant contamination of the HI recombination lines from the broad line region of the AGN. Therefore the intrinsic flux ratio of these lines is no longer due to case B in HII regions and so would provide inaccurate extinction values.

\subsubsection{Molecular Gas (H$_2$)}
\begin{figure}
	\includegraphics[width=\columnwidth]{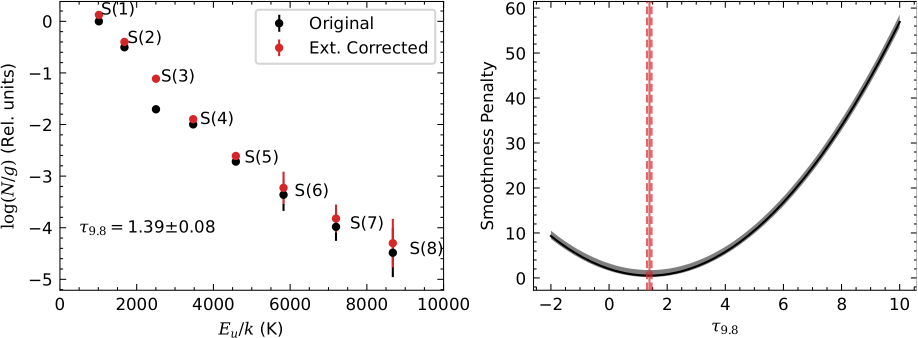}
    \caption{Inferring the extinction of H$_2$, in this case for NGC 3256 SF4, by varying the extinction correction and measuring the smoothness of the resultant rotation diagram. \textbf{Left:} Rotation diagram for H$_2$ for the observed fluxes (black) and the extinction corrected fluxes (red). Note the change for the S(3) transition. \textbf{Right:} Penalty factor for the smoothness of the resulting rotation diagram given an extinction correction with optical depth $\tau_{9.8}$. The best fit value is given by the minimum of this curve which is marked with the solid vertical red line. The vertical dashed red liens show the 1$\sigma$ error range.}
    \label{fig:RotDiagram} 
\end{figure}

The mid-infrared is populated with numerous H$_2$ lines tracing the molecular gas. In particular the S(1) to S(8) transitions between 5 $\mu$m and 28 $\mu$m. In \citet{Donnan2023b} and \citet{Hernandez2023}, it was noted that the S(3) J=5 transition is extremely sensitive to extinction as its central wavelength at 9.665 $\mu$m places it within the trough of the 9.8 $\mu$m silicate absorption \citep[see also][]{Pereira-Santaella2014}. In the referenced works, this line was largely ignored as a result, however this fact also makes it ideal for predicting level of extinction affecting the molecular gas. 
We do this by first constructing a rotation diagram where we plot the population of each energy level, $\log(N/g)$, against the energy of the upper level, $E_u/k$, as shown in the left panel of Fig. \ref{fig:RotDiagram}. This is for a typical star forming region (NGC 3256 SF4), where the fluxes used have not been corrected for extinction (black points). As a result, the S(3) transition appears lower than expected. We can therefore estimate the extinction by varying the correction factor until this dip no longer exists. 

We do this by varying the optical depth of a single screen of dust and calculating the extinction corrected rotation diagram for all the transitions. For each optical depth, we calculate the smoothness of the resulting rotation diagram using the sum of the Laplacian, which is the 1D version of equation (\ref{eqn:Laplace}), 
\begin{equation}
    P_{H_2} = \sum_{i} \left[ \log(N_{i-1}/g_{i-1}) - 2\log(N_{i}/g_{i}) + \log(N_{i+1}/g_{i+1}) \right]^2
\end{equation}
where, $\log(N_{i}/g_{i})$ is the population of transition $i$. The inferred extinction is that where the smoothness penalty, $P_{H_2}$ is minimised. This is demonstrated in the right panel of Fig. \ref{fig:RotDiagram} with the corresponding extinction corrected rotation diagram in the left panel in red. As before we can perform this for each MCMC sample to get an uncertainty on the inferred H$_2$ extinction. 

The gradient of the rotation diagram is related to the temperature of the gas, where steeper curves result from cooler gas as the higher energy transitions are relatively less populated. Typically one will fit multiple ($\sim $ 2 or 3) linear components, each with a different temperature to fit the curve \citep[e.g.][]{Rigopoulou2002, Pereira-Santaella2022} or assume some power law distribution of temperature \citep{Togi2016, Pereira-Santaella2014}. To obtain an estimate without assuming a model, we calculate the median gradient of pairs of transitions to provide an average molecular gas temperature as traced by the mid-infrared. 

\subsubsection{PAHs}
\label{sec:PAHExt}
If an intrinsic ratio between two PAH bands is assumed, the extinction can be inferred by the same method as the Hydrogen recombination lines. This method was proposed by \citet{Hernan-Caballero2020} where they showed that the 12.7/11.3 PAH ratio is intrinsically constant for star-forming galaxies with the observed ratio depending only on the extinction.

We calculate the intrinsic 12.7/11.3 PAH ratio using the star-forming regions of NGC 3256 and NGC 7469 as a calibration sample. This step is required as the increased spectral resolution of MIRI MRS over Spitzer IRS, leads to a more resolved 12.7 PAH feature \citep[e.g.][]{Donnan2023b}. We measure an intrinsic 12.7/11.3 PAH ratio of $0.72 \pm 0.08$, after correcting for extinction using the average continuum extinction as given in Table \ref{tab:Results}. As we rely on an empirical calibration, the resulting measured PAH extinctions will largely follow that of the continuum for the star-forming regions by construction. Therefore the measured PAH extinction is not an completely independent measure but is valuable as a comparison between objects.

Taking the samples of the intrinsic ratio, we vary the extinction until the observed 12.7/11.3 PAH ratio is reproduced for each MCMC sample of the measured ratio, to obtain a posterior for the PAH extinction.

\subsection{Results}
\label{sec:Results}
We fit the model to each spectrum extracted as discussed in Section \ref{sec:DataReduction}. We fit the entire sample with each emissivity, first with the simple power law with $\beta=2$ and then with the physical emissivity profile which includes a mixture of silicates and graphite based on \citet{Li2001}. We found the physical emissivity profile provides a better fit in all cases. Moreover it is necessary to fit the silicates seen in emission for the nucleus of NGC 7469 \citep[e.g][]{Alonso-Herrero2020, Garcia-Bernete2022b}. We compare the choice of emissivity in detail in Section \ref{sec:EmissivityDiscussion}. The results hereafter are with the physical emissivity, as it provides a better fit and is better motivated from a theoretical perspective.

We show the best fit dust distributions for all the nuclei and a representative star-forming region in Fig. \ref{fig:Results}. The remaining star-forming regions are shown in Appendix \ref{fig:Results2}. We use the physical emissivity as we found this to provide a better fit to the data and is more physically motivated.

We find the shape of the dust distribution, $\Psi \left ( T, \tau_{9.8} \right )$, to be largely consistent between the various star-forming regions. In particular it shows a relatively low dust temperature ($\sim$ 90 - 100 K) that is lightly obscured ($\tau_{9.8} \sim 0.2$), resulting in a steep dust continuum with little to no silicate absorption. There additionally appears to be a faint, obscured hot dust component with $\tau_{9.8}>1$. A few of the star-forming regions, NGC 7469 SF1, SF3 and SF4, show a stronger gradient, with a stronger hot dust component resulting in moderate silicate emission. Comparing with the other estimates of extinction, the molecular gas appears the most obscured, even more obscured than the dust continuum. We discuss possible reasons for this in Section \ref{sec:SFDiscussion}.

As expected, the nuclei generally show more complex dust distribution, where multiple dust components contribute to the observed spectrum. We discuss each individually in detail in Section \ref{sec:Nuclei}.


We display the inferred properties of each fitted spectrum in Table \ref{tab:Results}. The temperature, $T$, and extinction, $\tau_{9.8}$, are given for the total continuum and individual components of the dust distribution for those cases as described in Section \ref{sec:decomp}. To obtain this, we sum over one of the axes of the dust distribution to collapse the 2D distribution into a 1D distribution either of temperature, $T$, and extinction, $\tau_{9.8}$ (by summing over the other axis). We then determine the median and 16$\%$/84$\%$ upper and lower bounds of the 1D distributions. These values reflect the shape/spread of $\Psi(T, \tau_{9.8})$ rather than the statistical uncertainties, which are negligible by comparison.

We additionally show the measured extinction from the other tracers as outlined in Section \ref{sec:EstofExt}. For the molecular gas we also show the measured gas temperature.

For NGC 7469 Nuc and IIZw96 SW, the 12.7/11.3 PAH ratio is not well constrained, and therefore cannot measure the extinction accurately.

We note that the uncertainties on the inferred properties are dominated by the priors on the intrinsic PAH ratio as the statistical uncertainties are very small due to the high signal-to-noise of the data. This is reflected in the errors on the stellar extinction. These are likely underestimated as we have not folded in errors based on the assumption of the chosen stellar population templates, which for example, assumed solar metallicity.

\begin{table*}
\centering
  \caption{Inferred properties from each spectrum}
  \label{tab:Results}
    \def\arraystretch{1.5}
    \setlength{\tabcolsep}{2pt}
    \begin{threeparttable}
\resizebox{\textwidth}{!}{%
  \begin{tabular}{cccccccccccccc}
  
    \hline 
\multicolumn{1}{c}{Spectrum} &  \multicolumn{2}{c}{Total Cont.} &  \multicolumn{2}{c}{SF Cont.} & \multicolumn{2}{c}{AGN Warm} & \multicolumn{2}{c}{AGN Hot} & \multicolumn{1}{c}{HI} & \multicolumn{2}{c}{H$_2$} & \multicolumn{1}{c}{PAHs} & \multicolumn{1}{c}{Stellar}  \\
\cmidrule(r){2-3}
\cmidrule(r){4-5}
\cmidrule(r){6-7}
\cmidrule(r){8-9}
\cmidrule(r){10-10}
\cmidrule(r){11-12}
\cmidrule(r){12-12}
\cmidrule(r){13-13}
\cmidrule(r){14-14}

    &  $\tau_{9.8}$ & $T$ & $\tau_{9.8}$ & $T$ & $\tau_{9.8}$ & $T$ & $\tau_{9.8}$ & $T$ & $\tau_{9.8}$& $\tau_{9.8}$& $\bar{T}$&$\tau_{9.8}$ &$\tau_{9.8}$ \\
     & & K & & K & & K & & K &   & &K &  & \\
        \hline
NGC 7469 SF1&$0.49^{+1.2}_{-0.38}$ & $111.0^{+120.0}_{-50.0}$ &   - & - & - & - & - & - &0.56$^{+0.17}_{-0.18}$  &0.86$^{+0.05}_{-0.074}$ &716.0$^{+100.0}_{-100.0}$ &0.88$^{+0.27}_{-0.23}$ &0.3767$^{+0.0033}_{-0.0031}$ \\
NGC 7469 SF2&$0.43^{+1.6}_{-0.34}$ & $100.0^{+79.0}_{-41.0}$ &   - & - & - & - & - & - &1.09$^{+0.13}_{-0.14}$  &0.72$^{+0.084}_{-0.072}$ &664.0$^{+69.0}_{-74.0}$ &0.56$^{+0.28}_{-0.24}$ &0.3513$^{+0.0039}_{-0.0039}$ \\
NGC 7469 SF3&$0.40^{+1.2}_{-0.3}$ & $104.0^{+66.0}_{-40.0}$ &- & - & - & - & - & - &1.56$^{+0.29}_{-0.28}$  &1.11$^{+0.096}_{-0.11}$ &654.0$^{+54.0}_{-56.0}$ &0.54$^{+0.25}_{-0.24}$ &0.2436$^{+0.0036}_{-0.0037}$ \\
NGC 7469 SF4&$0.41^{+1.4}_{-0.31}$ & $100.0^{+65.0}_{-38.0}$ &   - & - & - & - & - & - &1.20$^{+0.20}_{-0.28}$  &0.87$^{+0.11}_{-0.13}$ &660.0$^{+76.0}_{-70.0}$ &0.55$^{+0.27}_{-0.21}$ &0.2401$^{+0.0027}_{-0.0027}$ \\
NGC 7469 SF5&$0.50^{+1.9}_{-0.4}$ & $93.0^{+75.0}_{-38.0}$ & - & - & - & - & - & - &0.55$^{+0.15}_{-0.21}$ &1.1$^{+0.11}_{-0.084}$ &629.0$^{+45.0}_{-46.0}$ &0.70$^{+0.24}_{-0.24}$ &0.2578$^{+0.0022}_{-0.003}$ \\
NGC 7469 SF6&$0.53^{+1.7}_{-0.41}$ & $105.0^{+87.0}_{-43.0}$ & - & - & - & - & - & - &0.55$^{+0.15}_{-0.16}$ &1.21$^{+0.084}_{-0.084}$ &654.0$^{+46.0}_{-42.0}$ &0.49$^{+0.25}_{-0.24}$ &0.3233$^{+0.0059}_{-0.0056}$ \\
NGC 7469 Nuc&$1.02^{+2.8}_{-0.89}$ & $154.0^{+62.0}_{-40.0}$ & - & - & $1.04^{+3.0}_{-0.92}$ & $149.0^{+47.0}_{-37.0}$ & $0.97^{+0.75}_{-0.48}$ &$634.0^{+390.0}_{-160.0}$ &  -  &0.34$^{+0.012}_{-0.024}$ &712.0$^{+5.9}_{-5.9}$ & - & - \\
NGC 3256 SF1&$0.33^{+1.5}_{-0.25}$ & $87.0^{+68.0}_{-33.0}$ &  - & - & - & - & - & - & 0.85$^{+0.28}_{-0.29}$  &1.26$^{+0.048}_{-0.06}$ &538.0$^{+92.0}_{-21.0}$ &0.15$^{+0.29}_{-0.23}$ &0.1862$^{+0.0035}_{-0.0033}$ \\
NGC 3256 SF2&$0.55^{+2.2}_{-0.44}$ & $90.0^{+68.0}_{-33.0}$ & - & - & - & - & - & - &1.10$^{+0.24}_{-0.22}$  &1.45$^{+0.048}_{-0.048}$ &684.0$^{+29.0}_{-34.0}$ &0.55$^{+0.24}_{-0.34}$ &0.4949$^{+0.0058}_{-0.0059}$ \\
NGC 3256 SF3&$0.38^{+1.6}_{-0.29}$ & $91.0^{+62.0}_{-33.0}$ &  - & - & - & - & - & - &0.72$^{+0.16}_{-0.15}$ &1.43$^{+0.036}_{-0.024}$ &722.0$^{+20.0}_{-24.0}$ &0.64$^{+0.26}_{-0.24}$ &0.3288$^{+0.0034}_{-0.0035}$ \\
NGC 3256 SF4&$0.41^{+1.7}_{-0.33}$ & $89.0^{+69.0}_{-36.0}$ & - & - & - & - & - & - &0.83$^{+0.19}_{-0.19}$  &1.39$^{+0.072}_{-0.084}$ &634.0$^{+58.0}_{-77.0}$ &0.39$^{+0.26}_{-0.24}$ &0.2575$^{+0.0052}_{-0.0052}$ \\
NGC 3256 SF5&$1.07^{+1.9}_{-0.81}$ & $116.0^{+84.0}_{-46.0}$ &   - & - & - & - & - & - &1.01$^{+0.060}_{-0.064}$  &1.63$^{+0.024}_{-0.024}$ &765.0$^{+14.0}_{-8.9}$ &0.40$^{+0.28}_{-0.22}$ &0.4173$^{+0.004}_{-0.0041}$ \\
NGC 3256 SF6&$0.47^{+1.7}_{-0.38}$ & $93.0^{+61.0}_{-33.0}$ &  - & - & - & - & - & - &0.81$^{+0.17}_{-0.17}$  &1.33$^{+0.048}_{-0.048}$ &658.0$^{+29.0}_{-52.0}$ &0.68$^{+0.25}_{-0.25}$  &0.3821$^{+0.0045}_{-0.0043}$ \\
NGC 3256 N&$1.07^{+1.3}_{-0.83}$ & $109.0^{+62.0}_{-29.0}$ &  - & - & - & - & - & - &1.35$^{+0.036}_{-0.036}$  &2.38$^{+0.012}_{-0.012}$ &436.0$^{+19.0}_{-3.2}$ &1.18$^{+0.26}_{-0.23}$ &0.2937$^{+0.0017}_{-0.0012}$ \\
NGC 3256 S&$4.66^{+6.4}_{-3.6}$ & $83.0^{+43.0}_{-27.0}$ & $0.66^{+0.61}_{-0.54}$ & $77.0^{+30.0}_{-22.0}$ & $6.1^{+6.4}_{-3.4}$ & $83.0^{+40.0}_{-27.0}$ & $6.6^{+2.7}_{-1.8}$ & $1274.0^{+230.0}_{-290.0}$ & 2.49$^{+0.040}_{-0.036}$  &2.24$^{+0.012}_{-0.024}$ &540.0$^{+38.0}_{-39.0}$ &2.07$^{+0.27}_{-0.22}$ &1.3971$^{+0.0018}_{-0.0018}$ \\
VV114 NE&$5.27^{+4.4}_{-2.6}$ & $93.0^{+68.0}_{-32.0}$ & $1.02^{+0.46}_{-0.88}$ & $64.0^{+23.0}_{-14.0}$ & $5.48^{+4.4}_{-2.5}$ & $95.0^{+68.0}_{-33.0}$ & - & - & 1.76$^{+0.058}_{-0.059}$  &1.98$^{+0.026}_{-0.012}$ &591.0$^{+11.0}_{-7.3}$ &2.03$^{+0.27}_{-0.23}$ &0.7016$^{+0.00056}_{-0.00056}$ \\
VV114 SW&$0.95^{+2.1}_{-0.81}$ & $84.0^{+100.0}_{-29.0}$ & $0.25^{+0.36}_{-0.17}$ & $71.0^{+27.0}_{-20.0}$ & $2.1^{+2.2}_{-1.0}$ & $93.0^{+66.0}_{-33.0}$ & $2.29^{+1.4}_{-0.83}$ & $732.0^{+260.0}_{-200.0}$ & 1.46$^{+0.038}_{-0.039}$  &1.51$^{+0.024}_{-0.024}$ &656.0$^{+92.0}_{-56.0}$ &1.47$^{+0.28}_{-0.22}$ &0.7309$^{+0.0046}_{-0.0043}$ \\
IIZw96 NE&$0.69^{+1.8}_{-0.55}$ & $72.0^{+70.0}_{-30.0}$ &  - & - & - & - & - & - &1.11$^{+0.14}_{-0.14}$ &0.69$^{+0.24}_{-0.31}$ &461.0$^{+120.0}_{-75.0}$ &1.49$^{+0.27}_{-0.24}$ &0.5045$^{+0.0037}_{-0.0038}$ \\
IIZw96 SW& $2.25^{+3.3}_{-1.7}$ & $91.0^{+58.0}_{-30.0}$ & $0.69^{+0.56}_{-0.54}$ & $76.0^{+35.0}_{-21.0}$ & $3.59^{+3.2}_{-1.5}$ & $104.0^{+60.0}_{-37.0}$ & $13.11^{+1.9}_{-9.5}$ & $774.0^{+340.0}_{-240.0}$ & 1.30$^{+0.020}_{-0.019}$ &2.08$^{+0.036}_{-0.024}$ &902.0$^{+9.2}_{-7.8}$ & -  &0.194$^{+0.0096}_{-0.0090}$\\
    \hline

  \end{tabular}}
  \end{threeparttable}
 \end{table*}

\begin{figure*}
	\includegraphics[width=\textwidth]{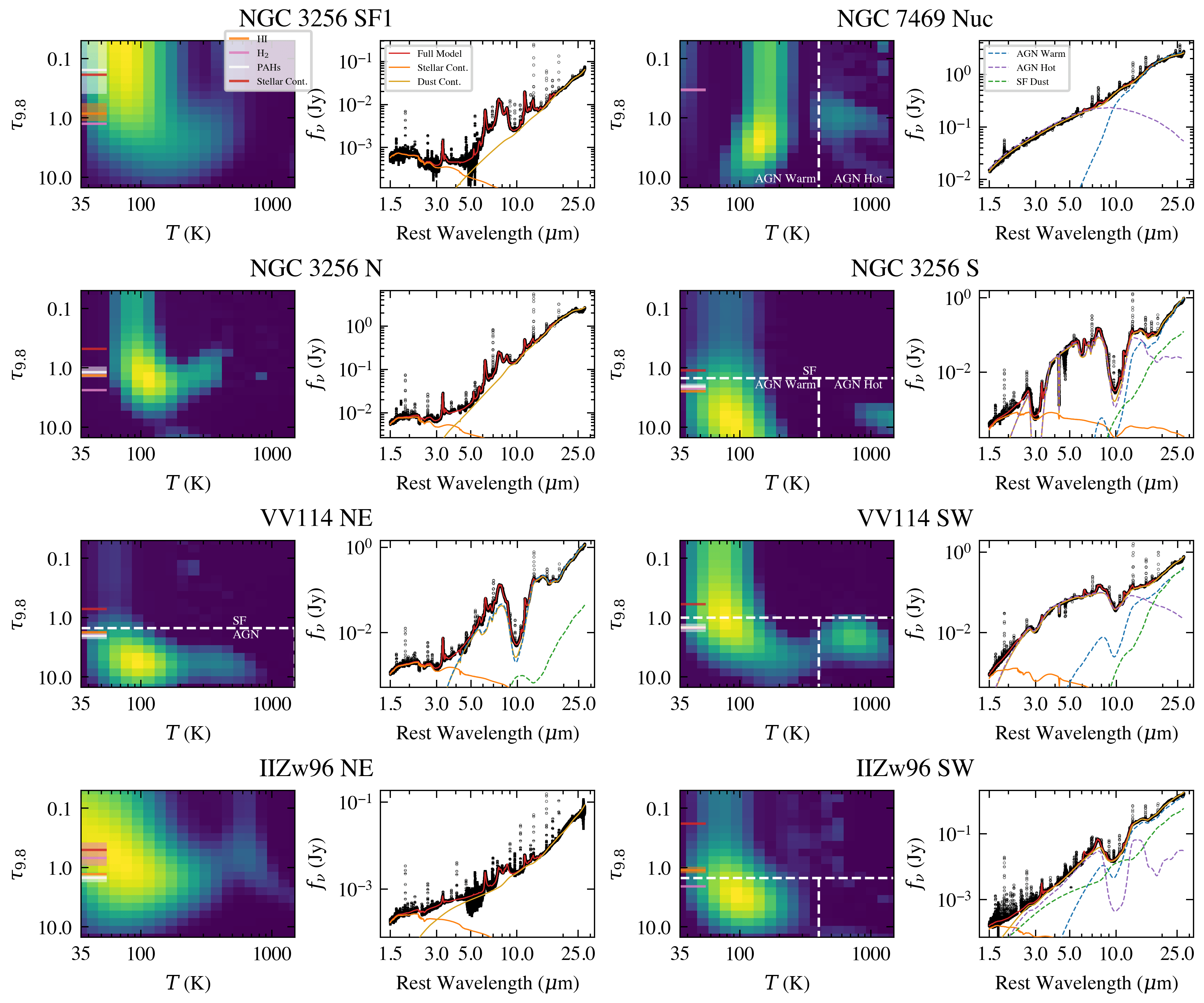}
    \caption{Fits to eight of the spectra in this work. We show a star-forming region (NGC 3256 SF1) and the 7 nuclei. Each panel shows the best fit dust distribution, $\Psi \left ( T, \tau_{9.8} \right ) $, on the left and the corresponding fit to the data in the right panel. The horizontal coloured lines show the extinction measurements for the molecular gas, PAHs, HI lines and the stellar continuum, as described in Section \ref{sec:EstofExt}. The dashed grey lines show a decomposition of the best fit distribution into star-forming and AGN torus + polar dust for those case where applied. The resulting continuum for each component is plotted in the right panels.}
    \label{fig:Results} 
\end{figure*}

\section{Discussion}
\label{sec:Discuss}

\subsection{Layers of Extinction}
\label{sec:SFDiscussion}
From Fig. \ref{fig:Results} and Fig. \ref{fig:Results2}, we
find that the molecular gas and HII regions are the most obscured while the PAHs and stellar continuum are less obscured. This is consistent with a picture where the molecular gas is the most buried within dust clouds which will collapse to form stars. The HII regions are similarly obscured, resulting from newly formed stars that still remain deeply buried within the star-forming region. 

The dust continuum appears highly mixed ranging from unobscured ($\tau_{9.8} \sim 0.2$) to obscured 
 ($\tau_{9.8} > 1$), with the most buried dust consistent with the molecular gas and HII regions. This suggests the dust exists throughout star-forming regions where a significant fraction of the flux is from the outer layers, where the dust emission is unobscured towards the line of sight of the observer. A significant portion of the heating of this unobscured dust may therefore originate from old stars that exist outwith the obscured molecular clouds where stars are actively forming. This is confirmed by the extinction from the stellar continuum, which at these wavelengths is dominated by evolved stars, and appears less buried than the more obscured molecular gas and HII regions. This is also true for the PAHs, where previous studies have shown that some excitation can be attributed to more evolved stellar populations \citep{Zhang2023b}, which is an important consideration when determining star-formation rates.

PAHs experience similar obscuration as the cooler dust continuum. This is however, largely by construction, as the intrinsic 12.7/11.3 ratio is empirically calibrated using the average extinction of the continuum (see Section \ref{sec:PAHExt}). However, one would expect the PAHs to be somewhat less obscured than the HII regions as they are known to exist in Photodisassociation Regions (PDRs), where the PAHs exist on the boundary between ionised and molecular gas \citep[e.g. See Fig. 1 of][]{Chown2023} and therefore the measured values (in Table \ref{tab:Results}) are not unexpected.

We find that the molecular gas temperature is generally high, likely due to these lines being sensitive to the warm molecular gas. If the S(0) transition was also included, the measured average temperature would likely be lower. We measure generally similar gas temperatures of $\sim 500 - 700$ K for all of the regions with the exception of IIZw96 SW, which shows a higher temperature of $\sim 900$ K. Interestingly, the known AGN, NGC 7469 Nuc and NGC 3256 S both show temperatures consistent with the star-forming regions. This is likely because the mid-IR transitions of H$_2$ originate from gas at temperatures around those shown in Table \ref{tab:Results} and so we are sensitive to molecular gas heating by stars rather than AGN. This has been found in previous studies \citep[e.g.][]{Rigopoulou2002} where AGN and star-forming dominated galaxies show similar molecular gas temperatures as traced by the mid-IR.

\subsection{Choice of Emissivity}
\label{sec:EmissivityDiscussion}
\begin{figure}
	\includegraphics[width=\columnwidth]{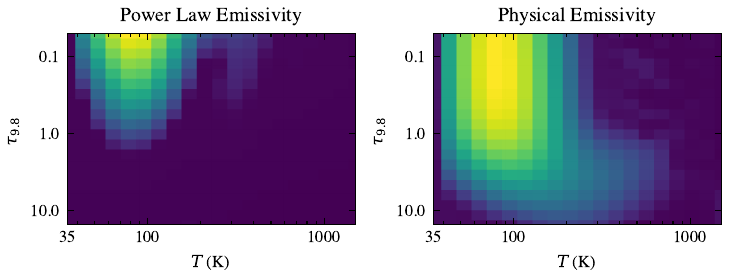}
    \caption{Comparison of the inferred dust distribution for a typical star-forming region (NGC 3256 SF3) depending on the chosen emissivity function as shown in Fig. \ref{fig:Emissivity}. The left is for $\epsilon_{\nu} \propto \lambda^{-2}$ while the right is a physically motivated emissivity function, including silicates, based on \citet{Li2001}.}
    \label{fig:EmmisivityComparison} 
\end{figure}

We found a more physically motivated emissivity, which includes silicates, provides a better fit to the data over a simple power law. In Fig. \ref{fig:EmmisivityComparison} we show a comparison of the inferred dust distribution for a star-forming region depending on the choice of emissivity.  

For the power law emissivity, the dust is almost completely unobscured and shows no sign of any differential extinction, as to create a smooth continuum with little silicate absorption requires close to zero extinction. By contrast the dust distribution for the physically motivated emissivity shows a greater average extinction compared to the power law emissivity, with evidence of differential extinction, with a more obscured hot dust component relative to the cooler dust. 

The latter case is physically more reasonable, as the dust extinction is more consistent with the other extinction tracers, where the buried hot dust is consistent with the obscured molecular gas where stars are actively forming. 
We therefore conclude that the physically motivated emissivity is closer to reality.

\subsection{Decomposing the Dust Distribution}
\label{sec:decomp}
For objects that show a complex dust distribution such as IIZw96 SW, we can decompose this into multiple components. For this example, there is a clear component of relatively unobscured cold dust that is consistent with the HI, H$_2$ and stellar continuum extinction measures. Additionally there is an obscured component with a gradient, where the hotter dust appears more obscured. This is potentially an AGN torus \citep{Garcia-Bernete2023b} 
and so if we split the dust distribution into two, we can extract individual continua for each component. 

As we find the majority of the dust distribution for star-forming regions lie at $\tau_{9.8}<1 \sim 1.5$, we define a cutoff around these values for NGC 7469 Nuc, VV114 NE, VV114 SW, IIZw96 SW and NGC 3256 S, as these show complex dust distributions with clear components. Firstly, we select dust from star-formation by using grid elements below $\tau_{9.8}<1.5$ for NGC 3256 S, VV114 NE and IIZw96 SW and $\tau_{9.8}<1.0$ for VV114 SW. We then note that NGC 7469 Nuc, NGC 3256 S and VV114 SW show a clear separate component in the $T, \tau_{9.8}$ space at high temperatures. We isolate this with a temperature cutoff of $T>400$ K. We also apply this to IIZw96 SW, which tentatively shows a separate component here.

The obscured, high temperature component is likely heated very close to the AGN, near the dust sublimation zone. This may be the inner rim of a torus or polar dust, where dusty grains are driven as a wind by radiation pressure from the AGN accretion disk \citep[e.g.][]{Honig2017, Garcia-Bernete2022d}.

We have presented a simple interpretation of the best fit dust distribution, but the reality may be more complex. For example if there is dust at a similar temperature and obscuration in both an AGN torus and star-forming regions, the two cannot be disentangled by a simple cutoff.

\subsection{Nature of the Individual Sources}
\label{sec:Nuclei}
From the best fit dust distributions and the extinction predicted from the different emission features, we can infer the nature of these sources in terms of layers of extinction. We briefly summarise our findings for each object here in the context of the literature. Beyond the scope of this work, the rich data of these targets allows much more analysis into the nature of these objects \citep[e.g.][]{Inami2022, Garcia-Bernete2022c, Donnan2023b, Rich2023, Armus2023, Lai2023, Zhang2023, Pereira-Santaella2023, Gonzalez-Alfonso2023, Garcia-Bernete2023b, Buiten2023}.

\subsubsection{NGC 7469}
\label{sec:NGC7469Dis}
We find the nuclear dust continuum to be dominated by the AGN, with relatively weak PAH emission and no significant contribution from stellar continuum, consistent with previous analysis of this target \citep[e.g.][]{Alonso-Herrero2020, Garcia-Bernete2022b, Armus2023}. The dust distribution shows a large spread of extinction with an average extinction of $\tau_{9.8} \sim 1$ with a significant hot dust component with $T > 400$ K. Considering its classification as a type 1 AGN, it is expected to have clear lines of sight to very hot dust close to the sublimation radius at T $> 1000$ K. Interestingly, this dust is moderately obscured which may suggest that there is some intervening dust that obscures the very hot inner regions of the AGN torus.

For three of the star-forming regions (see Section \ref{sec:DataReduction} and Fig. \ref{fig:Apertures}) we find the HII regions to be more obscured than the molecular gas, contrary to the other SF regions. Namely SF2, SF3 and SF4 show this behaviour and lie on the north-west section of the star-forming ring. Additionally, these regions show silicates in emission (along with SF1) due to a combination of low obscuration and higher dust temperatures. From \citet{Diaz-Santos2007}, these regions show slightly more evolved stellar populations and lack molecular gas (via CO (2-1)) compared to SF1, SF5 and SF6. This offers a potential explanation as to why the H$_2$ lines appear less obscured than the HII regions, where the relative lack of CO but presence of the mid-IR H$_2$ lines would suggest a high molecular gas temperature.

For NGC 7469 SF1, the comparative lack of cool obscuring dust, resulting in the silicate emission, may be related to the nuclear bar, as this region is placed on the boundary between the bar and the star-forming ring.

\subsubsection{VV 114}
\label{sec:VV114Dis}
VV 114 NE shows a significant highly obscured dust continuum appearing more obscured than the PAHs, molecular gas or  HII regions. This is consistent with initial analysis of this target in \citet{Donnan2023b} where there is likely an extremely buried AGN considering its obscuration and compact nature (<100 pc). Additionally, the ro-vibration molecular bands of CO and H$_2$O, show a high excitation temperature \citep{Buiten2023}. From the dust distribution, high temperatures are reached up to $\sim$ 800 K but unlike the SW nucleus, there is no clear isolated hot dust component.

VV 114 SW also shows a strong continuum but it is significantly less obscured than the NE nucleus. There is a significant isolated hot dust component that dominates the continuum between 2-10 $\mu$m.
This is consistent with the claim that this is an AGN by \citet{Rich2023}, due to its low 3.3 $\mu$m and 6.2 $\mu$m PAH equivalent widths, where the hot dust continuum dominates over the emission features. From the decomposition, the warm AGN component shows a gradient where the obscuration increases with temperature. This may indicate a high covering factor smooth torus. This picture can explain the lack of high ionisation potential lines \citep{Donnan2023b, Rich2023} that would be expected for an AGN with the observed level of obscuration.  

An alternative explanation is that the isolated hot dust component is a result of the shock front that has passed through the region \citep[e.g.][]{Saito2017, Donnan2023b} and that the nucleus is simply an obscured star-forming region. This may explain the lack of high-ionisation lines, and the low excitation temperature of the ro-vibration molecular bands \citep{Buiten2023}. However, similar to the NE core, the source is very compact (< 100 pc in the near-IR) and so reaching such a high infrared surface brightness is difficult without an AGN.

\subsubsection{NGC 3256}
\label{sec:NGC3256Dis}
NGC 3256S hosts a low luminosity, Compton Thick AGN \citep[e.g.][]{Ohyama2015} which has a collimated molecular outflow \citep[][]{Sakamoto2014, Emonts2014, Pereira-Santaella2023} detected through CO, HCN, HCO+ and H$_2$ transitions. From the dust distribution we see a clear signature of a buried continuum source which shows a gradient of extinction, with the extinction increasing with temperature in the warm component. This suggests a smooth torus rather than a clumpy one, as the hot dust is buried and not leaking through low density lines of sight. In addition, the dust distribution shows an isolated hot dust component that is also highly obscured. This hot dust, as shown in Fig. \ref{fig:Results}, dominates the continuum between 2 - 10 $\mu$m.

The stellar continuum, PAHs and HI extinction show a similar pattern to the star-forming regions and VV114 NE, where these trace different phases of the circumnuclear star-formation. However, the H$_2$ appears less obscured as it is likely tracing the molecular outflow dominating the H$_2$ emission over circumnuclear star-forming molecular clouds. 

The northern nucleus shows a dust distribution that is hotter with higher extinction than the star-forming regions but does not show any significant hot component that would be consistent with AGN heating. This is consistent with the literature, that this is a star-forming nucleus \citep{Lira2008}. 

Interestingly, the star-forming region to the southwest, SF5, shows a higher obscuration and dust temperature unlike the other star-forming regions as well as deeper near-IR ices. This may suggest an additional heating source such as supernovae and/or X-ray binaries. This is not unexpected considering LIRGs are believed to have high supernovae rates (many per year) owing to their high star-formation rates. Multiple supernovae have been detected in NGC 3256 \citep[e.g.][]{Kankare2018}, however none match this specific source. 

With the exception of SF5, all the star-forming regions in NGC 3256 are remarkably similar, with all showing the same pattern of obscuration where the stellar continuum, PAHs, HII regions and molecular gas exist from least obscured to most obscured respectively. The star-forming regions in NGC 3256 appear slightly cooler than those in NGC 7469 and therefore do not show any silicate emission. 

A detailed analysis of the PAH properties, comparing to theoretical spectra, will provide further insight (\textcolor{blue}{Rigopoulou et al. in prep.}).

\subsubsection{IIZw96}
\label{sec:IIZw96Dis}
The spectrum of the SW nuclei of IIZw96 is highly unusual \citep[][]{Inami2022, Garcia-Bernete2023b}, with very weak PAH features and a plethora of emission features in the near-infrared (< 3 $\mu$m). Indeed, \citet{Shipley2016} noted this object as an outlier in the PAH vs SFR correlation, due to its weak PAH emission as probed by Spitzer.

From the dust distribution we find evidence for a cold, buried continuum source, which shows a gradient, which would again suggest a smooth obscuring structure. The star-forming component is strong, which dilutes this obscured source, filling up the silicate feature. Additionally, the star-forming component is hot, leading to silicate emission at $9.8$ $\mu$m, consistent with the higher molecular gas temperature of $\sim900$ K. 

The $\sim 4.6 \mu$m CO band appears in emission rather than absorption \citep{Garcia-Bernete2023b}. 
This is the case for NGC 3256 S, where the nucleus shows CO absorption while the outflow shows emission \citep{Pereira-Santaella2023}. We tentatively detect a very hot dust component in the dust distribution and thus contributing to the continuum at $\sim$ 3 $\mu$m. Unlike NGC 3256 S and VV114 SW, this is not as clearly isolated from the warm component and so the temperature/extinction is not well constrained, however this component dominates the continuum in the near-IR band. The values for the extinction and temperature for this hot component in Table \ref{tab:Results} are consequently very uncertain. In particular the extinction, as the other dust components dominate in the silicate bands.

The NE nuclei appears to be a relatively unobscured star-forming region, however the PAH features appear a-typical. Most notably is the extremely strong 8.6 $\mu$m feature. Additionally, the molecular gas appears less obscured than the HII regions and PAHs.  
\subsection{Effective Extinction Curve}
For each object, the extinction correction factor, defined as the ratio of the observed to the intrinsic spectrum (equation (\ref{eqn:ExtCor})), varies depending on the shape of the best fit dust distribution, $\Psi\left(T, \tau_{9.8}\right)$. In general this dust distribution is not a screen and therefore cannot be converted into an optical depth as $\frac{f_{\nu}}{f_{\nu}^{\textrm{Int}}} \neq e^{ -\tau_{9.8} \tau(\lambda) }$. However, to compare the inferred dust correction factors with known extinction curves from the literature, we calculate an ``effective'' extinction curve by inverting this equation. This can be thought as ``locally'' assuming a screen at each wavelength and stitching together an extinction curve. It is worth noting that we are not deriving a new extinction law, as in general $\frac{f_{\nu}}{f_{\nu}^{\textrm{Int}}} \neq e^{ -\tau_{9.8} \tau(\lambda) }$ and therefore the variation we see between targets is purely due to the shape of $\Psi\left(T, \tau_{9.8}\right)$. In this work we have assumed a fixed law of that given in \citet{Donnan2023b}, which governs the intrinsic properties of the dust grains.

\begin{figure}
	\includegraphics[width=\columnwidth]{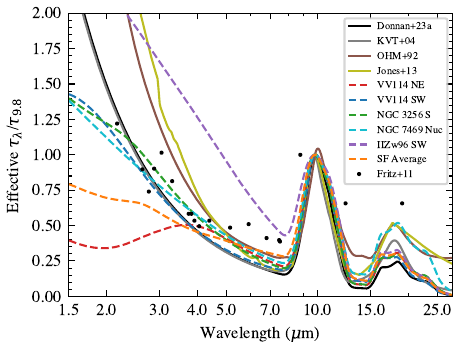}
    \caption{Comparison of the inferred extinction factor (eqaution (\ref{eqn:ExtCor})) for each target, converted to an ``effective'' optical depth, normalised at 9.8 $\mu$m. We show the targets in Fig. \ref{fig:PAHComps} as well as the average for the star-forming regions ing NGC 3256 and NGC 7469. We compare to the extinction law used in this work from \citet{Donnan2023b} and others from the literature, namely \citet{Ossenkopf1992, Kemper2004, Fritz2011, Jones2013}.}
    \label{fig:ExtComp} 
\end{figure}

Fig. \ref{fig:ExtComp} shows the ``effective'' optical depth normalised at 9.8 $\mu$m for a variety of our targets. 

All of the targets show a shallower gradient in the near-infrared compared to a single screen from any of the extinction laws. This highlights the effect of differential extinction, where the mid-infrared traces more obscured activity than the near-infrared.
This is particularly true for VV114 NE, where we observe a turnover at $\lambda < 4 \mu$m, where the relatively less obscured stellar continuum dominates over the highly buried dust continuum. IIZw96 also shows an unusual shape, where the silicate feature is ``filled up'' by the hot dust emission and thus the shorter wavelength portion appears at a relatively higher optical depth compared tro the silicate band.

Extinction curves with flatter optical depths at $\sim$ 5 $\mu$m, have been observed in the literature (see Fig. 4 of \citet{Draine2003} or Fig. 2 of \citet{Li2015}), deviating from the expected power law. Differential extinction offers a potential explanation, where the mid-infrared traces more obscured dust while the near-infrared traces less obscured stellar-continuum, flattening the ``effective'' extinction curve. When the near-infrared is dominated by buried hot dust instead, such as for NGC 3256 S, the ``effective'' extinction curve becomes steeper, better resembling a power law.

The ratio of the 9.8 $\mu$m and 18 $\mu$m silicate bands also varies between targets, depending on the relative obscuration and temperature of different dust components. This can relate to the clumpiness of AGN tori \citep[e.g.][]{Hatziminaoglou2015, Martinez-Peredes2020, Garcia-Bernete2023}.

\section{Conclusions}
We have presented a model to reproduce the joint NIRSpec and MIRI spectra between 1.5 - 28 $\mu$m for a variety of LIRGs. Our main findings are

\begin{itemize}
    \item By testing with simulated data, we show that the model is able to reproduce the dust distribution and thus extinction correction factor for input data with and without differential extinction, where in the latter case, the traditional screen/mixed approaches fail, highlighting the need for differential extinction to model the spectra of obscured (U)LIRGs.
    
    \item We find evidence for differential extinction from fitting the spectra of LIRGs where star-forming regions show dust distribution dominated by relatively unobscured cool dust with some buried hot dust.

    \item We find a deeply buried continuum component in the dust distribution for VV114 NE, NGC 3256 S and IIZw96 SW, with the latter two showing a hot isolated dust component at T$\sim 1000$ K, likely due to AGN heating. Similarly, VV114 SW shows a significant hot component, either due to AGN heating or a shock front passing through as a result of the galaxy-galaxy merger.

    \item We compare the dust extinction with that of the stellar continuum, PAHs, HII regions (HI lines) and molecular gas (H$_2$) lines. For star-forming regions we find that the molecular gas, and HII regions are the most obscured consistent with the hottest dust. The stellar continuum and PAHs are significantly less obscured, consistent with the cooler dust.

    \item By constructing an ``effective'' extinction curve, we find the near-infrared to be much shallower than the mid-infrared which is indicative of differential extinction, where the near-infrared traces less obscured emission than the mid-infrared. This can explain flatter extinction curves in the literature.

\end{itemize}
We have shown that by modelling the distribution of dust we are able to reproduce the spectrum for a variety of objects from star-forming regions to highly obscured AGN. Such detailed modelling is imperative considering the high frequency of dusty galaxies being observed with JWST.

\section*{Acknowledgements}
The authors are grateful to the DD-ERS team (Program 1328, PI: Lee Armus \& Aaron Evans) for the observing program with a zero–exclusive–access period.
FRD acknowledges support from STFC through grant ST/W507726/1. DR and IGB acknowledge support from STFC through grant ST/S000488/1 and ST/W000903/1. AAH acknowledges support from grant PID2021-124665NB-I00 funded by MCIN/AEI/10.13039/501100011033 and by ERDF A way of making Europe.

\section*{Data Availability}
All the data used in this work is publicly available as part of DD-ERS Program 1328, downloadable from the \href{https://mast.stsci.edu/portal/Mashup/Clients/Mast/Portal.html}{MAST archive}.



\bibliographystyle{mnras}
\bibliography{References} 




\appendix

\section{Example Dust Geometries}
\begin{figure}
	\includegraphics[width=\columnwidth]{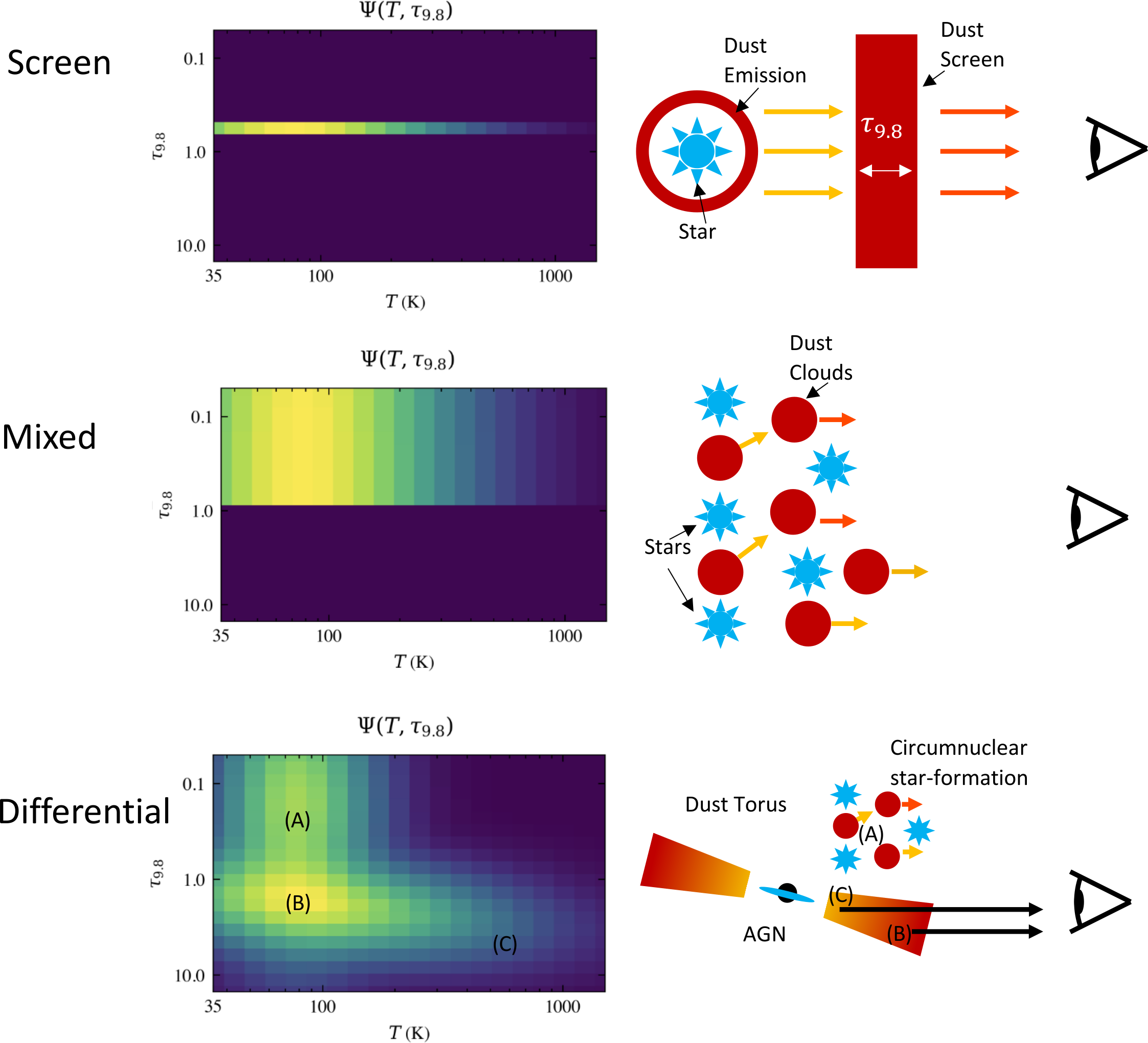}
    \caption{Diagrams demonstrating three different dust geometries discussed in this work. The left panels show the dust distribution, $\Psi \left ( T, \tau_{9.8} \right )$ corresponding to the geometry in the right panels. In each case the observer is on the right. The differential case is just one example where differential extinction is present. }
    \label{fig:Cartoons} 
\end{figure}

In Fig \ref{fig:Cartoons}, we show three diagrams demonstrating possible dust geometries we have described in this work. 

\section{PAH Features}
We show in Table \ref{tab:PAHs}, the parameters governing the Drude profiles used to model the PAH emission (equations (\ref{eqn:Drude}) and (\ref{eqn:Asymm})).

\begin{table}
\centering
  \caption{PAH Feature Parameters of equations (\ref{eqn:Drude},\ref{eqn:Asymm})}
  \label{tab:PAHs}
    \def\arraystretch{1.2}
    \setlength{\tabcolsep}{10pt}
    \begin{threeparttable}
  \begin{tabular}{ccc}
  
    \hline

     $\lambda_0$ & $\gamma_0$ & $a$  \\
     
      $\mu$m& $\mu$m&   \\
    (1) & (2) & (3) \\
        \hline

3.29 & 0.04 & 0.52 \\
3.40 & 0.03 & -10.0 \\
3.47 & 0.10 & -0.80 \\
5.18 & 0.05 & 0.00 \\
5.24 & 0.10 & -3.00 \\
5.45 & 0.15 & 0.00 \\
5.53 & 0.10 & 0.00 \\
5.64 & 0.10 & 0.00 \\
5.70 & 0.10 & 0.00 \\
5.76 & 0.10 & 0.00 \\
5.87 & 0.15 & 0.00 \\
6.00 & 0.20 & 0.00 \\
6.20 & 0.15 & -6.00 \\
6.69 & 0.40 & 0.00 \\
7.10 & 0.40 & 0.00 \\
7.42 & 0.94 & 0.00 \\
7.55 & 0.30 & 0.00 \\
7.61 & 0.10 & 0.00 \\
7.82 & 0.40 & 0.00 \\
8.33 & 0.20 & 0.00 \\
8.50 & 0.20 & 0.00 \\
8.61 & 0.34 & 0.00 \\
10.60 & 0.10 & 0.00 \\
10.74 & 0.10 & 0.00 \\
11.00 & 0.10 & -1.10 \\
11.20 & 0.10 & 0.00 \\
11.26 & 0.30 & -10.0 \\
11.99 & 0.54 & 0.00 \\
12.60 & 0.50 & 0.00 \\
12.77 & 0.15 & 0.00 \\
13.15 & 0.50 & 0.00 \\
13.55 & 0.20 & -5.00 \\
14.04 & 0.20 & 0.00 \\
14.19 & 0.20 & -5.00 \\
15.90 & 0.32 & 0.00 \\
16.45 & 0.23 & 0.00 \\
17.04 & 1.11 & 0.00 \\
17.38 & 0.21 & 0.00 \\
    \hline

  \end{tabular}
\begin{tablenotes}
    \item[] Column (1): Initial central wavelength for the fit with range $\pm 0.05\mu$m. Column (2): Initial FWHM allowed to vary within $+10\% $ and $-60\%$. Column (3): Initial asymmetry parameter $a$, allowed to vary within $+50\% $ and $-50\%$. Those with $a=0.00$ are fixed to a symmetric profile.
    \item[]
  \end{tablenotes}
  \end{threeparttable}
 \end{table}

\section{IIZw96 SW Contamination Correction}
\begin{figure}
	\includegraphics[width=\columnwidth]{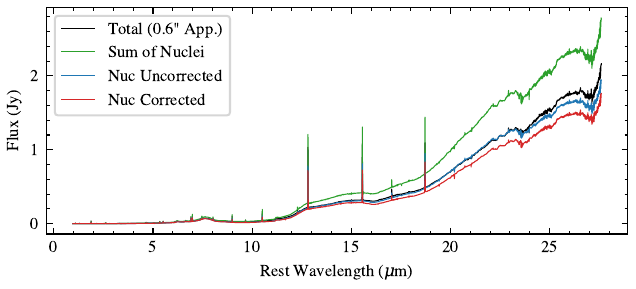}
    \caption{Correction of IIZw96 SW spectrum for contamination from nearby nucleus. The black line shows the total spectrum of the two nuclei from a 0.6" aperture, while the green shows the sum of the individual nuclei which displays an excess flux due to contamination. The blue spectrum shows the individual nucleus of interest which once corrected for contamination, returns the spectrum shown in red. }
    \label{fig:IIZw96Contamination} 
\end{figure}

Fig. \ref{fig:IIZw96Contamination} shows the correction of the extracted spectrum of IIZw96 SW for contamination, due to the nearby secondary nucleus. This process is described in detail in Section \ref{sec:DataReduction}.

\label{app:IIZw96}

\section{Additional SF Regions}
\begin{figure*}
	\includegraphics[width=\textwidth]{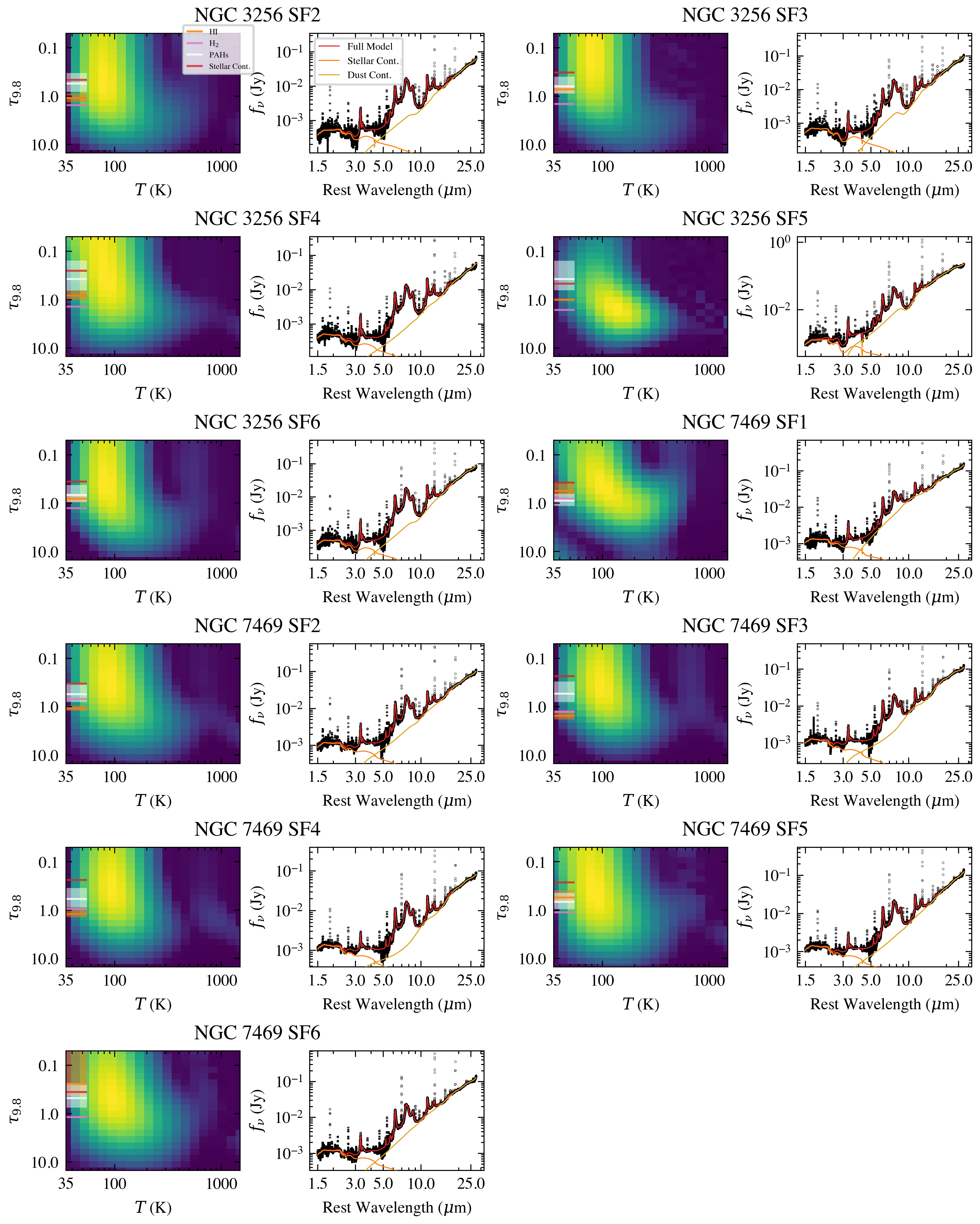}
    \caption{Same as Fig. \ref{fig:Results} but for the remaining spectra, all of which are star-forming regions. }
    \label{fig:Results2} 
\end{figure*}

Fig. \ref{fig:Results2} shows the remaining fits and inferred dust distributions for the spectra fitted in this work. These are the remaining star-forming regions from NGC 3256 and NGC 7469.     

\bsp	
\label{lastpage}
\end{document}